\documentclass[onecolumn]{autart}
\usepackage{cite}    % Enable this line and disable the
\usepackage[colorlinks,linkcolor=black,anchorcolor=red,citecolor=red]{hyperref}%

\usepackage{graphicx}           % Include figure files
\usepackage{dcolumn}            % Align table columns on decimal point
\usepackage{bm}                 % bold math
\usepackage{bbm}
\usepackage{graphics}           % for pdf, bitmapped graphics files
\usepackage{epsfig}             % for postscript graphics files
\usepackage{times}              % assumes new font selection scheme installed
\usepackage[fleqn]{amsmath}
\usepackage{amssymb}
\usepackage{extarrows}
\usepackage{amsfonts}
\usepackage{mathrsfs}
\usepackage{fancyhdr}
\usepackage{color}
\usepackage{subfigure}
\usepackage{enumerate}
\usepackage{epstopdf}
\usepackage{subfiles}
  \graphicspath{{images/}}
\usepackage [latin1]{inputenc} %
\usepackage{siunitx}

\usepackage{caption}
\usepackage{diagbox}

\allowdisplaybreaks[4] % [1] page breaks are allowed, but avoided if possible

\newtheorem{theorem}{Theorem}[section] %
\newtheorem{lemma}{Lemma}[section]

\newtheorem{definition}{Definition}[section]

\newtheorem{remark}{Remark}[section]

\newcommand{\diff}[1]{\operatorname{d}\!{#1}} % \diff{x}
\begin{document}

\begin{frontmatter}

\title{Power Tracking Control of Heterogeneous Populations of TCLs with Partially Measured States}

\author{Zhenhe Zhang$^{1}$}\ead{zhenhe.zhangzhu@polymtl.ca},
\author{Jun Zheng$^{1,2}$}\ead{zhengjun2014@aliyun.com},
\author{Guchuan Zhu$^{1}$}\ead{guchuan.zhu@polymtl.ca}

%%%%%%%%%%%%%%%%%%%%%%%%%%%%%%%%%%%%%%%%%%%%%%%
\address{$^{1}$Department of Electrical Engineering, Polytechnique Montr\'{e}al, P.O. Box 6079, Station Centre-Ville, Montreal, QC, Canada H3T 1J4\\
$^{2}${School of Mathematics, Southwest Jiaotong University,
        Chengdu 611756, Sichuan, China}
        }

%%%%%%%%%%%%%%%%%%%%%%%%%%%%%%%%%%%%%%%%%%%%%%%
\begin{keyword}
 Aggregate power tracking control, finite-time input-to-state stability, input-output linearization, partial differential equations, thermostatically controlled loads.
\end{keyword}
%%%%%%%%%%%%%%%%%%%%%%%%%%%%%%%%%%%%%%%%%%%%%%%
\begin{abstract}
 This paper presents a new aggregate power tracking control scheme for populations of thermostatically controlled loads (TCLs). The control design is performed in the framework of partial differential equations (PDEs) based on a late-lumping procedure without truncating the infinite-dimensional model describing the dynamics of the TCL population. An input-output linearization control scheme, which is independent of system parameters and uses only partial state measurement, is derived, and a sliding model-like control is applied to achieve finite-time input-to-state stability for tracking error dynamics. Such a control strategy can ensure robust performance in the presence of modeling uncertainties, while considerably reducing the communication burden in large scale distributed systems similar to that considered in the present work. A rigorous analysis of the closed-loop stability of the underlying PDE system was conducted, which guaranteed the validity of the developed control scheme. Simulation studies were performed while considering two TCL populations with a significant difference in their size, and the results show that the developed control scheme performs well in both cases, thereby confirming the effectiveness of the proposed solution.
\end{abstract}
\end{frontmatter}
%\tableofcontents
%%%%%%%%%%%%%%%%%%%%%%%%%%%%%%%%%%%%%%%%%%%%%%%%%%%%%%%%%%%%%%%%%%%%%%%%%%%%%%%%%%%%%%%%%%%%%%%%%
\section{Introduction}\label{Sec: Introduction}
In the context of today¡¯s smart grids, it is widely recognized that demand response (DR) programs have great potential in dealing with ongoing demands, while enhancing the energy efficiency and resilience of the power grid \cite{dms2,Siano2014,Haider2016,zhou2022exploiting}. As a promising demand-response enabled resource, thermostatically controlled loads (TCLs), such as air conditioners (ACs), space heating devices, refrigerators, and water heaters, are attracting increasing attention. Although a single TCL unit has very limited power regulation capability, ensembles of a large number of TCLs, when managed in an orderly and controllable manner, can have a significant impact on the entire power grid \cite{virtual2,battery,kong2020power}. It has been shown that a large TCL population can be managed to support demand response tasks, including peak load shaving and load following \cite{shaving2,following2,following3}, and to provide ancillary services, such as primary or secondary frequency controls \cite{primary1,primary3, primary4,webborn2019natural}.

The present work focuses on load tracking control, which allows the aggregate power of a TCL population to follow a desired consumption profile. The control design is based on a model of the dynamics of the TCL population described by partial differential equations (PDEs). Specifically, we consider a set of TCLs in which the dynamics of every individual device are modeled by a lumped stochastic hybrid system (SHS) operated through thermostat-based deadband control. The aggregate dynamics of such a TCL population can be modeled by two coupled Fokker-Planck equations (see, e.g, \cite{Malhame:1985,Callaway:2009,Zhao:2018}) describing the evolution of the probability distribution of TCLs in the ON and OFF states over the temperature. Note that the same form of PDE-based models can also be derived by assuming that the dynamics of individual TCLs are described by deterministic systems while considering population heterogeneity \cite{bashash2,Ghaffari:2015,moura2013modeling}.

Another widely adopted method to build the aggregate dynamical model of TCL populations is to divide a fixed range of temperatures into several segments, called state-bins, each of which is associated with the number of TCLs with their temperature fitting in this bin. The dynamics of state-bin transactions can be described by a Markov chain (see, e.g., \cite{prob3,bin1,Liu2016MPC,mahdavi2017model,Song2019_3}) or state queue (see, e.g., \cite{queue1,queue2}), which leads to finite-dimensional state-space models. It is worth noting that discretizing a PDE with respect to (w.r.t.) the space variable (temperature) also leads to a finite-dimensional state-space model. However, as the considered Fokker-Planck equation is a semi-linear time-varying PDE, its discretization results in a finite-dimensional nonlinear time-varying system. Consequently, a model described by the linear time invariant (LTI) system, which is the most used state-bin model in the existing literature, may be equivalent to that derived from PDEs only locally around particular equilibrium points and operational conditions (e.g., temperature set-point, ambient temperature, deadband), even with a variety of extensions. Therefore, the PDE provides a more generic framework for modeling the aggregate dynamics of TCL populations, which allows handling nonlinearity, time-varying operational conditions, and parametric uncertainties with often very simple control algorithms. However, the PDE control system design procedure generally involves more complex mathematical analysis and is more challenging.

The main objective of TCL population control is to manipulate the total power consumption of the entire population, which can be achieved by changing the temperature set-point, moving the deadband, or interfering with the probability distributions of the TCLs via forced switches (see, e.g., \cite{primary1,bashash2,Callaway:2009,mahdavi2017model,Totu:2017,ZLZL:2019}). Because a TCL population usually contains a large number of units that may spread over a large geographical area, only decentralized or distributed schemes are applicable control strategies. In fact, a remarkable amount of work on the control of TCL populations has been reported in the  literature, and the majority of the proposed solutions are based on lumped models by applying optimization theory and optimal control techniques, in particular model predictive control (see, e.g, \cite{primary1,primary4,prob3,bin1,Liu2016MPC,mahdavi2017model,Song2019_3,queue1,queue2,bashash2,Totu:2017}). It should be noted that, owing the nature of the considered problem, control schemes requiring the state measurement of the entire population in real-time are practically infeasible (see, e.g., \cite{dms1} and the references therein). This problem can be addressed using state observers \cite{primary4,Moura:CDC2013}. Nevertheless, it is still very challenging to assess the performance of model-based state estimation algorithms because it depends heavily on the accuracy of the system parameters.

The load tracking control algorithm developed in the present work is a decentralized scheme in which the rates for set-point temperature adjustment generated by a central unit are broadcast to the TCLs over the population. Emphasis is placed on solving issues arising in practical applications, particularly communication restrictions and modeling uncertainties for large scale TCL populations. The control system design is carried out in the framework of PDE-based modeling and control techniques. It should be noted that the two basic paradigms in PDE control system design and implementation, namely early-lumping and late-lumping procedures, have all been applied to the control of the coupled Fokker-Planck equations associated with TCL populations. The early-lumping method discretizes the underlying PDEs to obtain a lumped model, and then applies the techniques for finite-dimensional control system design \cite{bashash2,Callaway:2009,Ghaffari:2015, moura2013modeling, Totu:2017}. In contrast, with the late-lumping method, the controller is designed using the PDE model and then discretized for implementation \cite{Ghanavati:2018,ZLZL:2019}. A significant advantage of the late-lumping method is that it can preserve the essential properties of the PDE model and no approximation is required in the control design. However, some issues remain open. More specifically, the schemes developed in \cite{ZLZL:2019} and \cite{Ghanavati:2018} are based on input-output linearization by state feedback control, which may incur a communication burden. In addition, these control schemes require an accurate knowledge of the system parameters, for example, the diffusion coefficient in Fokker-Planck equations, which are not easy to determine from both theoretical and practical viewpoints considering the nature of the problem under investigation. Finally, although taking a weighted power load as the system output proposed in \cite{ZLZL:2019} can avoid the controllability issue introduced by the use of the total power load of the in-band TCLs as the system output in \cite{Ghanavati:2018}, such a choice lacks physical interpretation and is unsuitable for practical operation.

In this paper, we developed a new control algorithm based on the input-output linearization technique, which results in a system composed of finite-dimensional input-output dynamics and infinite-dimensional internal dynamics. The control design amounts then to finding a robust closed-loop control law that stabilizes the finite-dimensional input-output dynamics while guaranteeing the stability of the infinite-dimensional internal dynamics. Specifically:
\begin{itemize}
\item A new system output for power tracking control is proposed that can guarantee the controllability of the input-output dynamics.
\item A linearization control law, which is independent of system parameters, e.g., the diffusion coefficient, while requiring only knowledge of the states of TCLs near the deadband boundaries, is derived.
\item To tackle modeling uncertainties while making the control scheme computationally tractable, a sliding model-like tracking control scheme that can achieve finite-time input-to-state stability (FTISS)\cite{ftiss,Hong:2010}, is designed.
\item The non-negativeness of the solution to the Fokker-Planck equations under the developed control law and other properties required to ensure closed-loop stability are rigorously validated.
\end{itemize}

The main contribution of the present work lies in the simplicity, scalability, and applicability of the control strategy developed under a generic framework.
In addition, it is worth noting that as the developed control algorithm requires only measuring the state of the TCLs on the end-points of the deadband, TCLs need to notify their state only when switching occurs. Because the cyclic rate of the TCLs is much slower than the controller sampling rate, the communication burden can be significantly reduced. Obviously, it is very difficult for state feedback control schemes based on lumped aggregate models to achieve such features, which is critical for practical implementations.

%\subsection{Map out the paper}
The remainder of this paper is organized as follows. Section~\ref{Sec: notations} introduces the notations used in the study and preliminaries on FTISS. Section~\ref{models} presents the first-order equivalent thermal parameter (ETP) model for a single TCL unit and the coupled Fokker-Planck model for the aggregate dynamics of the TCL population. Section~\ref{Sec: Control design} presents the power tracking control design and closed-loop stability analysis. The experimental validation of the developed control strategy and the simulation results are reported in Section~\ref{Sec: experiment}, followed by concluding remarks in Section~\ref{Sec: Conclusion}. Finally, the proof of one of the main theoretical result is presented in the appendix.

\section{Notations and preliminaries}\label{Sec: notations}
\subsection{Notations}
Let $\mathbb{R} :=(-\infty,+\infty), $ $\mathbb{R}_{\geq 0}:=[0,+\infty)$, $\mathbb{R}_{> 0}:=(0,+\infty)$, and $\mathbb{R}_{\leq 0}:=(-\infty,0]$. Denote by $\partial_s f$ the derivative of the function $f$ {w.r.t.} argument $s$. Note that, for notation simplicity, we may omit the arguments of functions if there is no ambiguity.

By convention, we denote by $|\cdot|$ {the module of a function. For positive integers $m,n$  and a given  (open or closed)  domain $\Omega\subset \mathbb{R}^n$, let $ L^{\infty}(\Omega;\mathbb{R}^m):= \{ \phi:\Omega  \rightarrow\mathbb{R}^m|~\phi$ is measurable in $\Omega$ and satisfies $\text{ess sup}_{s\in \Omega }   |\phi(s)  | < +\infty \}$. For  $\phi\in L^{\infty}(\Omega;\mathbb{R}^m)$, the  norm of $\phi$ is defined by $ \|\phi\|_{L^{\infty}(\Omega)}:= \text{ess sup}_{s\in \Omega}   |\phi(s)| $. Let $ L^{\infty}_{loc}(\Omega;\mathbb{R}^m) :=\{\phi:\Omega  \rightarrow\mathbb{R}^m|~\phi\in  L^{\infty}(\Omega';\mathbb{R}^m)$ for any $\Omega'\subsetneqq\Omega\}$}

For  given   (open or closed) domains $\Omega_1,\Omega_2\subset \mathbb{R}^n$ and $\Omega_3\subset \mathbb{R} $, let $C \left(\Omega_1 ;\Omega_3  \right):= C^0\left(\Omega_1 ;\Omega_3  \right) :=\{\phi: \Omega_1 \rightarrow\Omega_3|~\phi$ is continuous  w.r.t.  its all augments in $\Omega_1\}$.  For positive integers $i,j$, let $C^i\left(\Omega_1;\Omega_3 \right):=\{\phi: \Omega_1 \rightarrow\Omega_3|~\phi$ has continuous derivatives   up to order $i$ w.r.t. its  all augments in $\Omega_1\}$, and $C^{i,j}\left(\Omega_1\times \Omega_2;\Omega_3 \right):=\{\phi: \Omega_1\times \Omega_2\rightarrow\Omega_3|~\phi$ has continuous derivatives   up to order $i$ w.r.t. its   augments in $\Omega_1$ and   up to order $j$ w.r.t. its   augments in $\Omega_2 \}$. In particular, if $\Omega_3=\mathbb{R}$, we denote $C\left(\Omega_1  \right):=C^0\left(\Omega_1;\mathbb{R}  \right)$ and $C^i\left(\Omega_1\right):=C^i\left(\Omega_1;\mathbb{R}  \right)$ for  $i>0 $.

As in \cite{ftiss} and \cite{Khalil:2002}, we define the following sets of comparison functions. Let $\mathcal{K}:=\{\vartheta : \mathbb{R}_{\geq 0} \rightarrow \mathbb{R}_{\geq 0}|\ \vartheta(0)=0,\vartheta$ is continuous, strictly increasing$\}$;
 $\mathcal{L}:=\{\vartheta : \mathbb{R}_{\geq 0}\rightarrow \mathbb{R}_{\geq 0}|\ \vartheta$ is continuous, strictly decreasing, $\lim_{s\rightarrow+\infty}\vartheta(s)=0\}$;
  $\mathcal{KL}:=\{\beta : \mathbb{R}_{\geq 0}\times \mathbb{R}_{\geq 0}\rightarrow \mathbb{R}_{\geq 0}|\ {\beta (\cdot,t)}\in \mathcal {K}, \forall t \in \mathbb{R}_{\geq 0}$, and $\beta(s,\cdot)\in \mathcal {L}, \forall s \in {\mathbb{R}_{+}}\}$;
    $\mathcal{K}_{\infty}:=\{\vartheta : \mathbb{R}_{\geq 0} \rightarrow \mathbb{R}_{\geq 0}| ~ \vartheta\in \mathcal{K}~\text{and}~\lim_{s\rightarrow+\infty}\vartheta(s)=+\infty\}$;
    $\mathcal{G}\mathcal{K}\mathcal{L}:=\{\beta:\mathbb{R}_{\geq 0}\times \mathbb{R}_{\ge0}\rightarrow \mathbb{R}_{\ge0}|~ \beta(\cdot,0)\in \mathcal{K} $, and for each fixed $s \in \mathbb{R}_{>0}$ there exists   $\widetilde{T}(s)\in \mathbb{R}_{\geq 0}$ such that $\beta(s,t)=0 $ for all $ t\geq\widetilde{T}(s)\}$.

\subsection{ Finite-time input-to-state stability of finite dimensional systems}
Consider the following nonlinear system
\begin{subequations}\label{eq: nonlinear system}
\begin{align}
  \dot{z}(t) =& f(z(t),d(t)),  \ \  \forall t\in \mathbb{R}_{\geq 0},\\
  z(0)= &z_0,
\end{align}
\end{subequations}
where $z:=[z_1,z_2,...,z_n]^T \in \mathbb{R}^n$ is the state, $z_0 \in \mathbb{R}^{n}$ is the initial datum, $d \in \mathcal{D}:=L^{\infty}_{\text{loc}}(\mathbb{R}_{\geq 0};\mathbb{R}^m)$ is the input (disturbance) to the system, $f: \mathbb{R}^n\times \mathbb{R}^m \rightarrow \mathbb{R}^n$ is a nonlinear function that is continuous {w.r.t. $(z,d)$}, ensures the forward existence of the
system solutions, at least locally,  and satisfies $ f(0, 0) = 0$, and $m\geq 1$ and $n\geq 1$ are integers.
\begin{definition} \label{Def.FTISS}
System~\eqref{eq: nonlinear system} is said to be finite-time input-to-state stable (FTISS) if {there exist functions $\vartheta\in \mathcal{K}$  and $\beta\in \mathcal{GKL}$ such that}  for any $x_{0}\in \mathbb{R}^{n}$
and  $d\in \mathcal{D}$ its trajectory satisfies
\begin{align}\label{eq: FTISS}
   |z(t)| \leq\beta( |z_0 |,t)+\vartheta(\|d\|_{{L^{\infty} (0,t)}}), \ \ \forall t \in \mathbb{R}_{\geq 0}.
\end{align}
\end{definition}
\begin{remark} Note that FTISS is defined in a similar way to the definition of  input-to-state stability (ISS) in  \cite[Chapter~4]{Khalil:2002} via the norm of $d$ over the interval $(0,t)$ rather than $(0,+\infty)$. Thus,  the FTISS  presented here is a refined notion of the one introduced in \cite{ftiss,Hong:2010}, where the second term in the right-hand side of \eqref{eq: FTISS} is under the form $\vartheta(\|d\|_{L^{\infty}(0,+\infty)})$, which describes the influence of the global bounds of $d$ instead of the bounds of $d$ over the finite time interval $(0,t)$.
\end{remark}
\begin{definition}
A continuously differentiable function $V:\mathbb{R}^{n}\rightarrow \mathbb{R}_{\ge0}$ is said to be an FTISS Lyapunov function for system~\eqref{eq: nonlinear system} if there exist functions $\mu_1,\mu_2\in\mathcal{K}_{\infty}$,   $\chi\in \mathcal{K}$ and constants $c>0$ and $\theta\in (0,1)$ such that for all $x\in \mathbb{R}^{n}$ and all $d\in \mathcal{D}$  it holds that
  %\begin{subequations}\label{eq: FTISS LF}
\begin{align*}
&{\mu_1({|x|})\leq  V(x)\leq \mu_2({|x|})},\\
&  |z| \geq   \chi (|d|)  \Rightarrow DV(z)\cdot f(z,d)\leq-cV^{\theta}(z),
\end{align*}
%\end{subequations}
where $DV(z):=\begin{bmatrix}\frac{\partial V}{\partial z_1},\ldots,\frac{\partial V}{\partial z_n}\end{bmatrix}$ .
\end{definition}
%\begin{remark}\cite{ftiss} mistake $V(x)=ln(1+|x|^2)$?
%\end{remark}

The following Lyapunov-like lemma   gives a sufficient condition for the FTISS.
\begin{lemma}\label{LemmaLyapunov}
System~\eqref{eq: nonlinear system} is FTISS if it admits a finite-time ISS Lyapunov function.
\end{lemma}
\begin{pf*}{Proof.} Setting $ \mathcal{V}:=\{z|V(z)\leq \mu_2(\chi(|d|))\}$ in the proof of \cite[Theorem~1(a)]{Hong:2010}, the lemma statement follows immediately. $\hfill\blacksquare$
\end{pf*}

 \section{Mathematical model and problem specification}\label{models}
\subsection{Dynamics of individual TCLs}
In the present work, we focus on modeling the population of residential air conditioners (ACs). While, its extension to other cooling and heating devices is straightforward. We consider the case where all ACs are operated by thermostats hence, every AC switches between the ON and OFF states whenever it reaches the prescribed lower or upper temperature bounds. For simplicity, we ignore the solar irradiation and internal heat gains and assume that the ACs operate at a fixed frequency. Then, the dynamics of the indoor temperature, denoted by $x$, for a representative load can be modeled by the following SHS (see, e.g., \cite{Malhame:1985,Callaway:2009,Totu:2017}):
\begin{align}\label{eqn:first-order}
 \diff{x}(t) = \dfrac{1}{CR}\left(x_a(t) - x(t) - s(t)RP\right)\diff{t} + \sigma \diff{w}(t),
\end{align}
where $x_a(t)$ is the ambient temperature, $R$, $C$, and $P$ are the thermal resistance, capacitance, and power, respectively, and $s(t)$ is the switching signal. In~\eqref{eqn:first-order}, $w(t)$ is a standard Wiener process, which, along with the parameter $\sigma$, represents modeling uncertainties, such as unaccounted heat loss or heat gain, parameter variations, and disturbances.

\begin{figure}[!htbp]
    \centering
    \captionsetup{justification=centering}
    \includegraphics[width=0.52\textwidth]{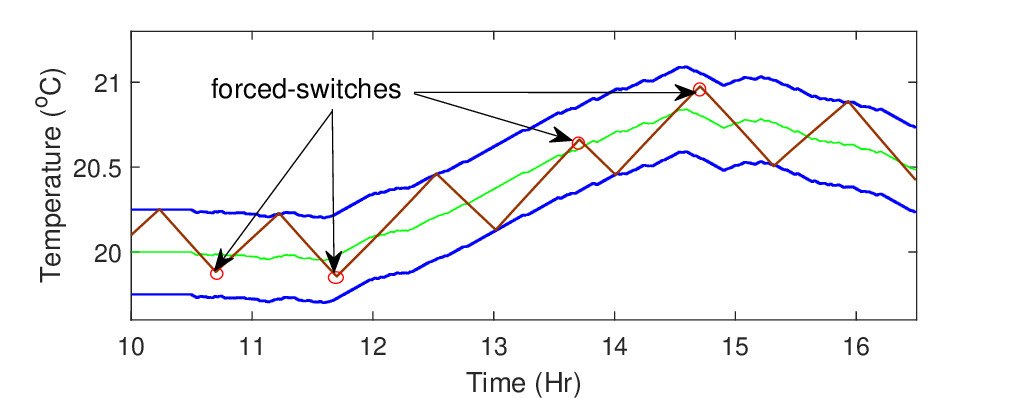}
    \caption{Hybrid thermostat-based deadband control scheme.}
    \label{fig:forced-switch}
\end{figure}

For a thermostat-controlled AC, the switching signal $s(t)$ takes a binary value from $\{0,1\}$, representing the OFF and ON states. We consider a hybrid control scheme, as shown in Fig.~\ref{fig:forced-switch}, in which the device always switches at the endpoints of the deadband. In addition, forced switches at any moment, denoted by $r(t)$, may also occur to alert the probability distributions of the TCL population. Let $r(t)$ take a binary value from $\{0,1\}$, with 1 representing the occurrence of switching and 0 otherwise. Letting $\underline{x}$ and $\overline{x}$ be the prescribed lower and upper temperature bounds, respectively, the deadband control for an AC can then be expressed as
\begin{align*}%\label{eq: sw}
  s(t) =
  \begin{cases}
    1, & \hbox{if~} x \geq \overline{x}; \\
    0, & \hbox{if~} x \leq \underline{x};\\
    (s(t^-)\wedge r(t)) + (s(t^-)\vee r(t)), & \hbox{otherwise};
  \end{cases}
\end{align*}
where ``$+$'' is the one-bit binary addition with overflow. In addition, the notations $(\cdot)^-$ and $(\cdot)^+$ denote the left and right limits of the scalar variable, respectively. Note that different actions, such as random switches to avoid power demand oscillations due to synchronization within a TCL population, mechanisms for blocking the switches to protect the ACs, etc., can be integrated in the design of forced switching schemes.

  \subsection{Dynamics of aggregate TCL population}\label{section:aggre}
\begin{figure}[!htbp]
    \centering
    \captionsetup{justification=centering}
    \includegraphics[width=0.52\textwidth]{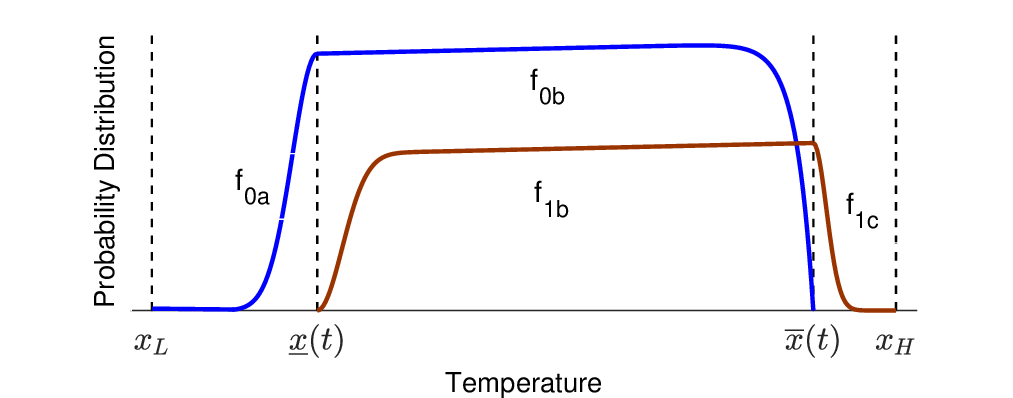}
    \caption{Illustration of probability density functions of a TCL population at a given time.}
    \label{fig: probability distribution}
\end{figure}

As mentioned previously, the dynamics of an aggregate TCL population can be characterized by the evolution of the distributions of the TCLs over temperature. When the number of TCLs in the population tends to be infinite, this population can be modeled as a continuum whose temperature distribution is governed by the coupled Fokker-Planck equations \cite{Malhame:1985,moura2013modeling,Ghaffari:2015, Callaway:2009}. Specifically, we denote by $f_{1}(x,t)$ and $f_{0}(x,t)$ the probability density functions (PDFs) of the TCLs in the ON and OFF states at temperature $x$ and time $t$, respectively. As illustrated in Fig.~\ref{fig: probability distribution}, we assume that all the loads are confined in a fixed temperature range $(x_L, x_H)$ along all possible operations, where $x_L$ and $x_H$ are constants, which is a reasonable assumption for practical application. Moreover, owing to the nature of thermostat-based control, there must be that $f_{1}(x,t) = 0$ for all $x \leq \underline{x}$ and $t\in\mathbb{R}_{>0}$, and that $f_{0}(x,t) = 0$ for all $x \geq \overline{x}$ and $t\in\mathbb{R}_{>0}$. Therefore, we can divide the range $(x_L, x_H)$ into three segments:
\begin{equation*}
  I_a:=(x_L,\underline{x}),I_b:=(\underline{x},\overline{x}),I_c:=(\overline{x},x_H),
\end{equation*}
which will be used in the upcoming study.

Suppose that the dynamics of each load in the TCL population are described by \eqref{eqn:first-order}. Let further
\begin{align*}
\alpha_0(x,t) :=& \frac{1}{CR}\left(x_a(t) - x \right), \\
\alpha_1(x,t) :=& \frac{1}{CR}\left({x_a(t)}  - x  - RP\right).
\end{align*}
The evolutions of $f_{0}(x,t)$ and $f_{1}(x,t)$ are governed by the following coupled Fokker-Planck equations \cite{Malhame:1985,Callaway:2009,Totu:2017}:
\begin{subequations}\label{eq: pdes}
\begin{align}
\partial_{t}f_{0}= & {\partial_{x}}\!\left(\dfrac{\sigma^{2}}{2}{\partial_{x}f_{0}}-(\alpha_{0}-u)f_{0}\right) \ \text{in~}I_{a}\times\mathbb{R}_{>0},\label{Pa}\\
\partial_{t}f_{0}= & {\partial_{x}}\!\left(\dfrac{\sigma^{2}}{2}{\partial_{x}f_{0}}-(\alpha_{0}-u)f_{0}\right)-g(f_{0},f_{1}) \ \text{in~}I_{b}\times\mathbb{R}_{>0},\label{Pb1}\\
\partial_{t}f_{1}= & {\partial_{x}}\!\left(\dfrac{\sigma^{2}}{2}{\partial_{x}f_{1}}-(\alpha_{1}-u)f_{1}\right)+g(f_{0},f_{1}) \
\text{in~}I_{b}\times\mathbb{R}_{>0},\label{Pb2}\\
\partial_{t}f_{1}= & {\partial_{x}}\!\left(\dfrac{\sigma^{2}}{2}{\partial_{x}f_{1}}-(\alpha_{1}-u)f_{1}\right) \ \text{in~}I_{c}\times\mathbb{R}_{>0},\label{Pc}
\end{align}
\end{subequations}
where $g(f_0,f_1)$ represents the net probability flux due to the switches occurring over segment $I_b$, that is, the so-called forced switches. Hence, the signs of $g(f_0,f_1)$ in \eqref{Pb1} and \eqref{Pb2} should be opposite to each other, which implies a mass conservation property as claimed in Theorem~\ref{Prop. conservation} in Section~\ref{Sec: Properties of the governing PDEs}. Note that \eqref{Pb1} and \eqref{Pb2} have a general form compared to that given in \cite{Totu:2017} (see (19a) and (19b) of that paper), where an explicitly linear function $g(f_0,f_1)$ was used to model a switching rate control scheme.

Following \cite{Totu:2017}, we introduce the notation of probability flows $\mathcal {F}_i$. When there is no additional flux from the forced switches, i.e., $g = 0$, $\mathcal {F}_i$ is the integral of the probability fluxes $\partial_{t}f_{i}$ over the temperature ($x$-) coordinate:
\begin{equation*}
\mathcal {F}_i(x,t) :=\dfrac{\sigma^{2}}{2}\partial_{x}f_{i}(x,t) -(\alpha_{i} (x,t)-u(t))f_{i}(x,t) ,i=0,1.
\end{equation*}
The boundary conditions can then be written as
\begin{subequations}\label{eq: BC-pf}
\begin{align}
 \mathcal{F}_0(x_L^+,t)=& 0,\ \ \forall t\in\mathbb{R}_{>0},\\
 \mathcal{F}_0(\underline{x}^-,t)=&  \mathcal{F}_0(\underline{x}^+,t)+\mathcal{F}_1(\underline{x}^+,t),\ \ \forall t\in\mathbb{R}_{>0},\\
  f_0(\underline{x}^-,t)= &f_0(\underline{x}^+,t),\ \ \forall t\in\mathbb{R}_{>0},\label{BC0-continuity}\\
    f_0(\overline{x}^-,t)=&0,\ \ \forall t\in\mathbb{R}_{>0},\\
  f_1(\underline{x}^+,t)=& 0,\ \ \forall t\in\mathbb{R}_{>0},\\
       f_1(\overline{x}^-,t)= & f_1(\overline{x}^+,t),\ \ \forall t\in\mathbb{R}_{>0},\label{BC1-continuity}\\
    \mathcal{F}_1(\overline{x}^+,t)=&  \mathcal{F}_0(\overline{x}^-,t)+\mathcal{F}_1(\overline{x}^-,t), \ \ \forall t\in\mathbb{R}_{>0},\\
    \mathcal{F}_1(x_H^-,t)=&0,\ \ \forall t\in\mathbb{R}_{>0},\\
     \mathcal{F}_0(\underline{x}^-,t)>&  \mathcal{F}_0(\underline{x}^+,t),\ \ \forall t\in\mathbb{R}_{>0},\\
  \mathcal{F}_1(\overline{x}^+,t)<&\mathcal{F}_1(\overline{x}^-,t),\ \ \forall t\in\mathbb{R}_{>0}.
\end{align}
\end{subequations}
 %Note that \eqref{BC0-continuity} and \eqref{BC1-continuity} assure the continuity of $f_0$ in $(\underline{x},t)$ and  $f_1$ in $(\overline{x},t)$, respectively.
% In addition, it should hold that
% \begin{subequations}\label{eq: BC-pf-2}
%  \begin{align}
%  \mathcal{F}_0(\underline{x}^-,t)>&  \mathcal{F}_0(\underline{x}^+,t)+,\ \ \forall t\in\mathbb{R}_{>0},\\
%  \mathcal{F}_1(\overline{x}^+,t)<&\mathcal{F}_1(\overline{x}^-,t),\ \ \forall t\in\mathbb{R}_{>0}.
%  \end{align}
%\end{subequations}

The initial data of $f_0$ and $f_1$ defined over $\overline{I}_{a0}:= [x_L,\underline{x}(0)] ,\overline{I}_{b0}:= [\underline{x}(0),\overline{x}(0)]$, and $\overline{I}_{c0}:= [ \overline{x}(0),x_H]$ are given by
\begin{subequations}\label{eq: IC}
\begin{align}
 f_0(0,x) = &f_0^{a0}(x), \ \ \forall x\in \overline{I}_{a0}, \\
 f_0(0,x) = &f_0^{b0}(x), \ \ \forall x\in \overline{I}_{b0}, \\
 f_1(0,x) = &f_1^{b0}(x), \ \ \forall x\in \overline{I}_{b0}, \\
 f_1(0,x) = &f_1^{c0}(x), \ \ \forall x\in \overline{I}_{c0}.
\end{align}
\end{subequations}

The total power demand of the TCL population at time $t \in \mathbb{R}_{\geq 0}$ is given by
\begin{align}\label{desired}
   y_{\text{total}}(t) :=\frac{P}{\eta} \int_{\underline{x}(t)}^{x_{H}}f_1(x,t)\diff{x},
\end{align}
where $\eta$ is the load efficiency coefficient.

\begin{remark} We provide remarks on the boundary conditions presented in \eqref{eq: BC-pf}.
\begin{enumerate}
\item[(i)] For continuous functions  $\alpha_0,\alpha_1$, and $u$,  the boundary conditions in \eqref{eq: BC-pf} are equivalent to:
\begin{subequations}\label{eq: BC}
\begin{align}
\dfrac{\sigma^{2}}{2}{\partial_{x}f_{0}}(x_{L}^{+},t)=&(\alpha_{0}(x_{L}^{+},t)-u(t))f_{0}(x_{L}^{+},t)
 ,\label{BC1}\\
 {\partial_{x}f_{0}}(\underline{x}^{-},t)=& {\partial_{x}f_{0}}(\underline{x}^{+},t)+ {\partial_{x}f_{1}}(\underline{x}^{+},t),
 \label{BC2}\\
 f_0(\underline{x}^-,t)= &f_0(\underline{x}^+,t),\\
 f_{0}(\overline{x},t) =&0, \label{BC3}\\
 f_{1}(\underline{x},t) =&0, \label{BC4}\\
 f_1(\overline{x}^-,t)= & f_1(\overline{x}^+,t),\\
{\partial_{x}f_{1}}(\overline{x}^{+},t)=& {\partial_{x}f_{0}}(\overline{x}^{-},t)+{\partial_{x}f_{1}}(\overline{x}^{-},t) ,\label{BC5}\\
 \dfrac{\sigma^{2}}{2}{\partial_{x}f_{1}}(x_{H}^{-},t)=&(\alpha_{1}(x_{H}^{-},t)-u(t))f_{1}(x_{H}^{-},t) ,
 \label{BC6}\\
  \partial_{x}f_1(\underline{x}^+,t)>& 0,\label{BC8} \\
\partial_{x}f_0(\overline{x}^-,t)<&0.\label{BC7}
\end{align}
\end{subequations}

\item[(ii)]  It is worth noting that this set of boundary conditions (\eqref{eq: BC-pf} or \eqref{eq: BC}), with possible variations, is commonly used in the literature \cite{Malhame:1985,Callaway:2009,Totu:2017}, which captures the basic properties of the considered problem, for example, impenetrable wall reflections (\eqref{BC1} and \eqref{BC6}), absorbing actions due to thermostat switching (\eqref{BC3}) and \eqref{BC4}), and probability conservation at the boundaries of the deadband (\eqref{BC2} and \eqref{BC5}). Note  that because of the absorbing property and the continuity of the PDFs on the boundaries of the deadband, the conditions \eqref{BC2} and \eqref{BC5} remain the same as those originally derived in \cite{Malhame:1985}, even though the considered problem in the present work contains control actions.
\end{enumerate}
\end{remark}

\subsection{{Problem statement and basic assumptions}}\label{Sec: assumption}
 {In this work, we study the dynamics described by the PDE model \eqref{eq: pdes} under the boundary and initial conditions \eqref{eq: BC-pf} and \eqref{eq: IC}. Based on \eqref{desired}, a new output function will be defined and specified in Section~\ref{Sec: Control design}. With these dynamics, a continuous time controller that considers the convergence time and robustness is designed to {stabilize} the tracking process.}\par
In the sequel, we  assume that $x_a\in {C( \mathbb{R}_{\geq 0} )}$
$\underline{x},\overline{x}\in C^1(\mathbb{R}_{\geq 0};\mathbb{R}_{>0})$, and
 denote
\begin{align*}
%  I_a:=&(x_L,\underline{x}),I_b:=(\underline{x},\overline{x}),I_c:=(\overline{x},x_H),\\%, I_{ab}:= (x_L,\overline{x}),I_{bc}:=(\underline{x},x_H).
% \end{align*}
% and
%\begin{align*}
S_{ab}:=& \left({C^{2,1}}(I_{a}\times\mathbb{R}_{>0})\cap C(\overline{I}_{a}\times\mathbb{R}_{\geq0})\right) \cup\left({C^{2,1}}(I_{b}\times\mathbb{R}_{>0})\cap C(\overline{I}_{b}\times\mathbb{R}_{\geq0})\right),\\
S_{bc}:=&	\left({C^{2,1}}(I_{b}\times\mathbb{R}_{>0})\cap C(\overline{I}_{b}\times\mathbb{R}_{\geq0})\right) \cup\left({C^{2,1}}(I_{c}\times\mathbb{R}_{>0})\cap C(\overline{I}_{c}\times\mathbb{R}_{\geq0})\right).
\end{align*}

Based on the physical properties of the problem, we impose the following structural conditions and basic assumptions on the solution and control for the system:
\newline
$\bullet$ The function of net probability flux $g$ belongs to $   C^1(\mathbb{R} ^2;\mathbb{R})$ and satisfies
\begin{enumerate}
  \item[(G1)]  $g(0,\tau)\leq 0$ for all $\tau\in \mathbb{R} $;
  \item[(G2)]  $g(s,0)\geq 0$ for all $s\in \mathbb{R}$;
  \item[(G3)]  $|g_{s}(s,\tau)|+|g_{\tau}(s,\tau)|\leq 1$ for all $(s,\tau)\in\mathbb{R}^2$.
\end{enumerate}
$\bullet$  The pair of solution $(f_0,f_1)$ and the control $u$ satisfy
\begin{enumerate}
  %\item[(A1)] $\underline{x},\overline{x}\in C^1(\mathbb{R}_{\geq 0};\mathbb{R}_{>0})$;
  \item[(U)] $u\in C(\mathbb{R}_{\geq 0};\mathbb{R})$ such that $\dot{\underline{x}} =\dot{\overline{x}} =u $ in $ \mathbb{R}_{\geq 0}$;
      % \item[$\bullet$]   $ v _{0}\in C(\overline{I}_{av};\mathbb{R}_{\geq 0}),   w _{0}\in C(\overline{I}_{bc};\mathbb{R}_{\geq 0}) $;
  \item[(F1)] $ f_0^{a0}\in C(\overline{I}_{a0};\mathbb{R}_{\geq 0})$, $f_0^{b0}\in C(\overline{I}_{b0};\mathbb{R}_{\geq 0})$, $f_1^{b0}\in C(\overline{I}_{b0};\mathbb{R}_{\geq 0})$, $f_1^{c0}\in C(\overline{I}_{c0};\mathbb{R}_{\geq 0})$;
  \item[(F2)] $f_0\in S_{ab}$ and has derivatives $\partial_{x}f_0(x_L^+,t)$, $\partial_{x}f_0(\underline{x}^{\pm},t)$ and $\partial_{x}f_0 (\overline{x}^-,t)$ for any  fixed $t\in\mathbb{R}_{>0}$;
  \item[(F3)] $f_1\in S_{bc}$ and has derivatives $\partial_{x}f_1(x_H^-,t)$, $\partial_{x}f_1(\overline{x}^{\pm},t)$ and $\partial_{x}f_1 (\underline{x}^+,t)$ for any fixed $t\in\mathbb{R}_{>0}$.
  %\item[(A6)] For  any fixed $t\in\mathbb{R}_{>0}$,  the probability flows  $\mathcal{F}_i$ satisfies
%  \begin{align}
%  \mathcal{F}_0(\underline{x}^-,t)>&  \mathcal{F}_0(\underline{x}^+,t), \\
%  \mathcal{F}_1(\overline{x}^+,t)<&\mathcal{F}_1(\overline{x}^-,t),
%  \end{align}
%  or, equivalently,
%  $\partial_{x}f_0(\overline{x}^-,t)<0$ and  $\partial_{x}f_1(\underline{x}^+,t)> 0$ .
\end{enumerate}
\begin{remark}
It should be mentioned that for $f_0=0$ (or $f_1=0$), condition~(G1) (or (G2)) guarantees $-g(f_0,f_1)\geq 0$ (or $g(f_0,f_1)\geq 0$) in \eqref{Pb1} (or \eqref{Pb2}) . This indicates that forced switching, which generates additional fluxes, is only possible from the $f_1$~system into the $f_0$~system when $f_0$ is zero.

Condition~(G3) indicates that the change in the probability density of the additional flux cannot be too fast for practical applications. {This is in accordance with the suggestion in \cite{Totu:2017}.}

Condition~(F1) indicates that the initial data are assumed to be nonnegative and continuous over the given domains. Conditions (F2) and (F3) describe the regularity of the solutions at the endpoints of the given domains at any time $t$.
\end{remark}

\section{Control design and stability analysis}\label{Sec: Control design}
In this section, we design a feedback control to ensure that the output of the system \eqref{eq: pdes}-\eqref{eq: IC} tracks a reference power curve, and assess the stability of the error dynamics in the framework of FTISS theory. Moreover, we study the mass conservation and non-negativeness properties of the solutions to the considered system, which allows further clarification of the physical meanings of the mathematical model.

\subsection{Control design}\label{smc}
The control objective is to drive the power consumption of the population to track the desired regulation signal. To this end, we choose an output of the power tracking control scheme as
%\begin{align}\label{eqn:output}
\begin{align}\label{eqn:output}
y(t):=&y_{\text{total}}(t) + \frac{P}{\eta} \int_{\overline{x}(t)}^{x_{H}}f_1(x,t)\diff{x}  - \frac{P}{\eta} \int_{x_{L}}^{\underline{x}(t)}f_0(x,t)\diff{x},  \ \ t \in \mathbb{R}_{\geq 0}.
\end{align}
%\end{align}

It is worth noting that, as the probability flows of $f_0$ and $f_1$ always move towards the deadband, $y(t)$ defined in \eqref{eqn:output} converges to the aggregated power demand $y_{\text{total}}(t)$ in the steady state. The motivation to add two extra terms to $y_{\text{total}}(t)$ is to ensure the controllability of the input-output dynamics.
%This issue will be addressed in detail later in Section~\ref{Sec: Properties of the governing PDEs}.

The regulation of power consumption of the TCL population is achieved by moving the mass of the temperature distribution, and the control signal is chosen to be the set-point temperature variation rate $\dot{x}_{sp}$, which may induce a change in the probability flux \cite{Callaway:2009,Zhao:2018}. As we consider a control scheme with a fixed deadband width, denoted by $\delta_{0}$, we have $\overline{x}=x_{sp}-\frac{\delta_{0}}{2},\;\underline{x}=x_{sp}+\frac{\delta_{0}}{2}$. Thus, the actual control signal is given by $u(t):= \dot{x}_{sp} = \dot{\underline{x}}=\dot{\overline{x}}$.

Let $y_{d}: {\mathbb{R}_{\geq 0} \rightarrow  \mathbb{R}}$ be the desired power profile, which is sufficiently smooth, and define the power tracking error as
\begin{equation*}
  e(t):=y(t)-y_{d}(t).
\end{equation*}

In what follows, we introduce a nonlinear control law and derive the corresponding tracking error dynamics.

\begin{theorem}\label{Thm:error}
Consider the system given in \eqref{eq: pdes} and \eqref{eqn:output} under the boundary conditions in \eqref{eq: BC-pf} (or equivalently \eqref{eq: BC}). Let the control input be defined as
\begin{align}\label{ideal-controller}
 {u(t):=}\dfrac{k|e(t)|^{\gamma}\mathrm{sgn}(e(t))+\Phi(t)}{2\left(f_{1}(\overline{x},t)+f_{0}(\underline{x},t)\right)},
\end{align}
where $k\in \mathbb{R}_{>0}$ and $\gamma\in (0,1)$ are constants,  $\mathrm{sgn}(e)$ is the sign function defined by
\begin{align*}
\mathrm{sgn}(e) & :=\begin{cases}
-1, & e < 0,\\
0, & e= 0,\\
1, & e> 0,
\end{cases}
\end{align*}
and
\begin{align}\label{PhiT}
\Phi(t):=& -\frac{\eta}{P}\dot{y}_{d}(t).
\end{align}
Then, the power tracking error dynamics are given by
\begin{align}\label{error}
\dot{e}(t) = -\frac{P}{\eta}k|e(t)|^{\gamma}\mathrm{sgn}(e(t))  +\Gamma(t),
\end{align}
where
\begin{align}\label{GammaT}
\Gamma(t):= & \dfrac{P}{\eta}\left(\alpha_{1}(\overline{x},t)f_{1}(\overline{x},t)+\alpha_{0}(\underline{x},t)f_{0}(\underline{x},t)\right)
                -\dfrac{\sigma^{2}P}{2\eta}\left({\partial_{x}f_{1}(\underline{x}^{+},t)}
              +{\partial_{x}f_{1}(\overline{x}^{+},t)}\right)\nonumber\\
           &  -\dfrac{\sigma^{2}P}{2\eta}\left({\partial_{x}f_{0}(\underline{x}^{-},t)}
              +{\partial_{x}f_{0}(\overline{x}^{-},t)}\right) +\frac{P}{\eta}\int_{\underline{x}(t)}^{\overline{x}(t)}g(f_{0},f_{1})\diff{x}.
\end{align}
\end{theorem}

\begin{remark}
$\Gamma(t)$ defined in \eqref{GammaT} captures the terms depending on the diffusion coefficient or requiring instantaneous state measurements and will be treated as a disturbance thereafter. Moreover, the control law given in \eqref{ideal-controller} involves only the measurement of the states (probability distributions $f_{0}$ and $f_{1}$) on the end-points of the deadband ($\underline{x}$ and $\overline{x}$), which results in a control scheme with significantly reduced communication burden compared to control schemes that require full-state measurements.
\end{remark}

\begin{pf*}{Proof of Theorem~\ref{Thm:error}.}
Note that
\begin{align*}
\dot{e}(t)=&\dot{y}(t)-\dot{y}_{d}(t)\\=&\frac{\text{d}}{\text{d}t}
\left(\frac{P}{\eta}\int_{\underline{x}(t)}^{x_{H}}f_{1}(x,t)\diff{x}
+\frac{P}{\eta}\int_{\overline{x}(t)}^{x_{H}}f_{1}(x,t)\diff{x} -\frac{P}{\eta}\int_{x_{L}}^{\underline{x}(t)}f_{0}(x,t)\diff{x}\right)-\dot{y}_{d}(t)\\
=&\frac{P}{\eta}\frac{\text{d}}{\text{d}t}
\int_{\underline{x}(t)}^{x_{H}}f_{1}(x,t)\diff{x}+\frac{P}{\eta}\frac{\text{d}}{\text{d}t}\int_{\overline{x}(t)}^{x_{H}}f_{1}(x,t)\diff{x} -\frac{P}{\eta}\frac{\text{d}}{\text{d}t}\int_{x_{L}}^{\underline{x}(t)}f_{0}(x,t)\diff{x}-\dot{y}_{d}(t).
\end{align*}
Hence, we decompose the whole computation process into three steps.\\

\emph{Step~1:} Compute $\frac{\text{d}}{\text{d}t}\int_{\overline{x}(t)}^{x_{H}}f_{1}(x,t)\diff{x}$. It follows immediately from Leibniz's integral rule and \eqref{Pc} that
\begin{align*}
&\frac{\text{d}}{\text{d}t}\int_{\overline{x}(t)}^{x_{H}}f_{1}(x,t)\diff{x}\\
=&0-\dot{\overline{x}}(t)f_{1}(\overline{x},t)+\int_{\overline{x}(t)}^{x_{H}}\partial_{t}f_{1}(x,t)\diff{x}\\
=&-u(t)f_{1}(\overline{x},t) +\int_{\overline{x}(t)}^{x_{H}}{\partial_{x}}\!\left(\dfrac{\sigma^{2}}{2}{\partial_{x} f_{1}(x,t)}
-(\alpha_{1}(x,t)-u(t))f_{1}(x,t)\right)\diff{x}\\=&-u(t)f_{1}(\overline{x},t) +\left(\dfrac{\sigma^{2}}{2}{\partial_{x}f_{1}(x_{H}^{-},t)}-(\alpha_{1}(x_{H}^{-},t)-u(t))f_{1}(x_{H}^{-},t)\right) -\left(\dfrac{\sigma^{2}}{2}{\partial_{x}f_{1}(\overline{x}^{+},t)}-(\alpha_{1}(\overline{x},t)-u(t))f_{1}(\overline{x},t)\right).
\end{align*}

Using boundary condition \eqref{BC6}, it follows that
\begin{align}\label{Step1}
 \frac{\text{d}}{\text{d}t}\int_{\overline{x}(t)}^{x_{H}}f_{1}(x,t)\diff{x}  
 =&-2u(t)f_{1}(\overline{x},t)
   -\dfrac{\sigma^{2}}{2}{\partial_{x}f_{1}(\overline{x}^{+},t)}+\alpha_{1}(\overline{x},t)f_{1}(\overline{x},t).
\end{align}

\emph{Step~2:}  Compute $\frac{\text{d}}{\text{d}t}\int_{\underline{x}(t)}^{x_H}f_{1}(x,t)\diff{x}$. Since
\begin{align*}
 \frac{\text{d}}{\text{d}t}\int_{\underline{x}(t)}^{x_{H}}f_{1}(x,t)\diff{x} 
%=&\frac{\text{d}}{\text{d}t}\left(\int_{\underline{x}(t)}^{\overline{x}(t)}f_{1}(x,t)\diff{x}
%+\int_{\overline{x}(t)}^{x_{H}}f_{1}(x,t)\diff{x}\right)\\
=&\frac{\text{d}}{\text{d}t}\int_{\underline{x}(t)}^{\overline{x}(t)}f_{1}(x,t)\diff{x}
  +\frac{\text{d}}{\text{d}t}\int_{\overline{x}(t)}^{x_{H}}f_{1}(x,t)\diff{x},
\end{align*}
and $\frac{\text{d}}{\text{d}t}\int_{\overline{x}}^{x_{H}}f_{1}(x,t)\diff{x}$ is given by \eqref{Step1}, we only need to compute $\frac{\text{d}}{\text{d}t}\int_{\underline{x}(t)}^{\overline{x}(t)}f_{1}(x,t)\diff{x}$. It follows from \eqref{Pb2} and \eqref{BC4} that
\begin{align}
 \frac{\text{d}}{\text{d}t}\int_{\underline{x}(t)}^{\overline{x}(t)}f_{1}(x,t)\diff{x}  
%=&\dot{\overline{x}}(t)f_{1}(\overline{x},t)-\dot{\underline{x}}(t)f_{1}(\underline{x},t)
% +\int_{\underline{x}(t)}^{\overline{x}(t)}\partial_{t}f_{1}(x,t)\diff{x}\nonumber\\
%=&u(t)f_{1}(\overline{x},t)-0+\int_{\underline{x}(t)}^{\overline{x}(t)}\partial_{t}f_{1}(x,t)\diff{x}\nonumber\\
%=&u(t)f_{1}(\overline{x},t)+\int_{\underline{x}(t)}^{\overline{x}(t)}g(f_{0},f_{1})\diff{x}\\
%&+\!\int_{\underline{x}(t)}^{\overline{x}(t)}\!\!\!\left({\partial_{x}}\!\left(\dfrac{\sigma^{2}}{2}{\partial_{x}f_{1}(x,t)}-(\alpha_{1}(x,t)-u(t))f_{1}(x,t)
%\right)\!\!\right)\diff{x} \notag\\
%&+\int_{\underline{x}(t)}^{\overline{x}(t)}g(f_{0},f_{1})\diff{x} \\
 =&u(t)f_{1}(\overline{x},t)+\int_{\underline{x}(t)}^{\overline{x}(t)}g(f_{0},f_{1})\diff{x} +\int_{\underline{x}(t)}^{\overline{x}(t)}\!{\partial_{x}}\!\left(\dfrac{\sigma^{2}}{2}{\partial_{x}f_{1}(x,t)}
  -(\alpha_{1}(x,t)-u(t))f_{1}(x,t)\right)\diff{x} \nonumber\\
=&u(t)f_{1}(\overline{x},t)+\int_{\underline{x}(t)}^{\overline{x}(t)}g(f_{0},f_{1})\diff{x}  +\left(\dfrac{\sigma^{2}}{2}{\partial_{x}f_{1}(\overline{x}^{-},t)}-(\alpha_{1}(\overline{x}^{-},t)-u(t))f_{1}(\overline{x}^{-},t)\right) \nonumber\\
& -\left(\dfrac{\sigma^{2}}{2}{\partial_{x}f_{1}(\underline{x}^{+},t)}-(\alpha_{1}(\underline{x}^{+},t)-u(t))f_{1}(\underline{x}^{+},t)\right)\nonumber\\
=&u(t)f_{1}(\overline{x},t)+\dfrac{\sigma^{2}}{2}{\partial_{x}f_{1}(\overline{x}^{-},t)}-(\alpha_{1}(\overline{x},t)-u(t))f_{1}(\overline{x},t) -\dfrac{\sigma^{2}}{2}{\partial_{x}f_{1}(\underline{x}^{+},t)}\notag\\
&+\int_{\underline{x}(t)}^{\overline{x}(t)}g(f_{0},f_{1})\diff{x}\nonumber\\
=&2u(t)f_{1}(\overline{x},t)+\dfrac{\sigma^{2}}{2}{\partial_{x}f_{1}(\overline{x}^{-},t)}-\alpha_{1}(\overline{x},t)f_{1}(\overline{x},t) 
 -\dfrac{\sigma^{2}}{2}{\partial_{x}f_{1}(\underline{x}^{+},t)}+\int_{\underline{x}(t)}^{\overline{x}(t)}g(f_{0},f_{1})\diff{x}. \label{eq: Step2_1}
\end{align}
%Therefore,
%\begin{align*}
%&\frac{\text{d}}{\text{d}t}\int_{\underline{x}(t)}^{x_{H}}f_{1}(x,t)\diff{x}\\
%=&\frac{\text{d}}{\text{d}t}\int_{\underline{x}(t)}^{\overline{x}(t)}f_{1}(x,t)\diff{x}
%+\frac{\text{d}}{\text{d}t}\int_{\overline{x}(t)}^{x_{H}}f_{1}(x,t)\diff{x}\\
%=&2u(t)f_{1}(\overline{x},t)+\dfrac{\sigma^{2}}{2}{\partial_{x}f_{1}(\overline{x}^{-},t)}-\alpha_{1}(\overline{x},t)f_{1}(\overline{x},t)\\
%&-\dfrac{\sigma^{2}}{2}{\partial_{x}f_{1}(\underline{x}^{+},t)}
%+\int_{\underline{x}(t)}^{\overline{x}(t)}g(f_{0},f_{1})\diff{x}-2u(t)f_{1}(\overline{x},t)\\
%&-\dfrac{\sigma^{2}}{2}{\partial_{x}f_{1}(\overline{x}^{+},t)}+\alpha_{1}(\overline{x},t)f_{1}(\overline{x},t)\\
%=&\dfrac{\sigma^{2}}{2}{\partial_{x}f_{1}(\overline{x}^{-},t)}-\dfrac{\sigma^{2}}{2}{\partial_{x}f_{1}(\underline{x}^{+},t)}
%+\int_{\underline{x}(t)}^{\overline{x}(t)}g(f_{0},f_{1})\diff{x}\\
%& -\dfrac{\sigma^{2}}{2}{\partial_{x}f_{1}(\overline{x}^{+},t)} ~~~~(\text{by}~~\eqref{BC6})\\
%=&-\dfrac{\sigma^{2}}{2}{\partial_{x}f_{1}(\underline{x}^{+},t)}-\dfrac{\sigma^{2}}{2}{\partial_{x}f_{0}(\overline{x}^{-},t)}
%+\int_{\underline{x}(t)}^{\overline{x}(t)}g(f_{0},f_{1})\diff{x}.
%\end{align*}
Combining \eqref{Step1} and \eqref{eq: Step2_1} we obtain by \eqref{BC6}
\begin{align}\label{Step2}
 \frac{\text{d}}{\text{d}t}\int_{\underline{x}(t)}^{x_{H}}f_{1}(x,t)\diff{x} 
=&-\dfrac{\sigma^{2}}{2}{\partial_{x}f_{1}(\underline{x}^{+},t)}-\dfrac{\sigma^{2}}{2}{\partial_{x}f_{0}(\overline{x}^{-},t)}
 +\int_{\underline{x}(t)}^{\overline{x}(t)}g(f_{0},f_{1})\diff{x}.
\end{align}

\emph{Step~3:}  Compute $\frac{\text{d}}{\text{d}t}\int_{x_{L}}^{\underline{x}(t)}f_{0}(x,t)\diff{x}$. According to \eqref{Pa} and \eqref{BC1}, we have
\begin{align}
 \frac{\text{d}}{\text{d}t}\int_{x_{L}}^{\underline{x}(t)}f_{0}(x,t)\diff{x}  
%=&\dot{\underline{x}}(t)f_{0}(\underline{x},t)-0+\int_{x_{L}^{+}}^{\underline{x}(t)}\partial_{t}f_{0}(x,t)\diff{x}\nonumber\\
=&u(t)f_{0}(\underline{x},t)+\int_{x_{L}}^{\underline{x}(t)}\partial_{t}f_{0}(x,t)\diff{x}\nonumber\\
=&u(t)f_{0}(\underline{x},t)+\int_{x_{L}}^{\underline{x}(t)}{\partial_{x}}\!\left(\dfrac{\sigma^{2}}{2}{\partial_{x}f_{0}}
 -(\alpha_{0}-u)f_{0}\right)\diff{x}\nonumber\\
=&u(t)f_{0}(\underline{x},t)+\left(\dfrac{\sigma^{2}}{2}{\partial_{x}f_{0}(\underline{x}^{-},t)}-(\alpha_{0}(\underline{x},t)
 -u(t))f_{0}(\underline{x},t)\right)\nonumber\\
&-\left(\dfrac{\sigma^{2}}{2}{\partial_{x}f_{0}(x_{L}^{+},t)}-(\alpha_{0}(x_{L}^{+},t)-u(t))f_{0}(x_{L}^{+},t)\right)\nonumber\\
%=&2u(t)f_{0}(\underline{x},t)+\dfrac{\sigma^{2}}{2}{\partial_{x}f_{0}(\underline{x}^{-},t)}-\alpha_{0}(\underline{x},t)f_{0}(\underline{x},t)-0\\
=&2u(t)f_{0}(\underline{x},t)+\dfrac{\sigma^{2}}{2}{\partial_{x}f_{0}(\underline{x}^{-},t)}-\alpha_{0}(\underline{x},t)f_{0}(\underline{x},t).
\label{Step3}
\end{align}
%Consequently,
%\begin{align}\label{Step3}
%&\frac{\text{d}}{\text{d}t}\int_{x_{L}}^{\underline{x}(t)}f_{0}(x,t)\diff{x}\nonumber \\
%=&2u(t)f_{0}(\underline{x},t)+\dfrac{\sigma^{2}}{2}{\partial_{x}f_{0}(\underline{x}^{-},t)}-\alpha_{0}(\underline{x},t)f_{0}(\underline{x},t).
%\end{align}
Finally, by combining \eqref{Step1}, \eqref{Step2}, and \eqref{Step3}, we obtain:
\begin{align*}
 \dot{e}(t) 
=&\frac{P}{\eta}\left(-\dfrac{\sigma^{2}}{2}{\partial_{x}f_{1}(\underline{x}^{+},t)}-\dfrac{\sigma^{2}}{2}{\partial_{x}f_{0}(\overline{x}^{-},t)}
\right) +\frac{P}{\eta}\left(-2u(t)f_{1}(\overline{x},t)
-\dfrac{\sigma^{2}}{2}{\partial_{x}f_{1}(\overline{x}^{+},t)}+\alpha_{1}(\overline{x},t)f_{1}(\overline{x},t)\right)\\
&-\frac{P}{\eta}\left(2u(t)f_{0}(\underline{x},t)+\dfrac{\sigma^{2}}{2}{\partial_{x}f_{0}(\underline{x}^{-},t)}
-\alpha_{0}(\underline{x},t)f_{0}(\underline{x},t)\right) -\dot{y_{d}}(t)+\frac{P}{\eta} \int_{\underline{x}(t)}^{\overline{x}(t)}g(f_{0},f_{1})\diff{x}\\
=&-\frac{2P}{\eta}u(t)\left(f_{1}(\overline{x},t)+f_{0}(\underline{x},t)\right)-\dfrac{\sigma^{2}P}{2\eta}{\partial_{x}f_{1}(\underline{x}^{+},t)} -\dfrac{\sigma^{2}P}{2\eta}{\partial_{x}f_{1}(\overline{x}^{+},t)}-\dfrac{\sigma^{2}P}{2\eta}{\partial_{x}f_{0}(\underline{x}^{-},t)}\\
&-\dfrac{\sigma^{2}P}{2\eta}{\partial_{x}f_{0}(\overline{x}^{-},t)}+\frac{P}{\eta}\int_{\underline{x}(t)}^{\overline{x}(t)}g(f_{0},f_{1})\diff{x} +\frac{P}{\eta}\alpha_{1}(\overline{x},t)f_{1}(\overline{x},t)+\frac{P}{\eta}\alpha_{0}(\underline{x},t)f_{0}(\underline{x},t)-\dot{y}_{d}(t).
\end{align*}
The error dynamics can then be expressed as
\begin{equation*}
  \dot{e}(t)=-\frac{2P}{\eta}u(t)\left(f_{1}(\overline{x},t)+f_{0}(\underline{x},t)\right)+ \frac{P}{\eta}\Phi(t) + \Gamma(t).
\end{equation*}

Let
\begin{align*}%\label{pre-controller}
  u(t) :=\dfrac{v(t)+\Phi(t)}{2\left(f_{1}(\overline{x},t)+f_{0}(\underline{x},t)\right)},
\end{align*}
where $v(t)$ is an auxiliary control input, then
\begin{align}\label{linear error dyn}
\dot{e}(t) = -\frac{P}{\eta}v(t)+\Gamma(t).
\end{align}
Considering an auxiliary control of the form:
\begin{align}\label{auxilary}
  v(t):=k|e(t)|^{\gamma}\mathrm{sgn}(e(t)),
\end{align}
the tracking error dynamics in the closed loop are then given by \eqref{error}.
$\hfill\blacksquare$
\end{pf*}

\begin{remark}
Note that for the given  initial data (see (F1)), it can be shown that the term $f_{1}(\overline{x},t)+f_{0}(\underline{x},t)$ is strictly positive (see Theorem~\ref{Prop-nonnegative}~(iii) in Section~\ref{Sec: FTISS}). Therefore, the control signal $u$, given in~\eqref{ideal-controller} is well-defined. In addition, $u$ is continuous due to the fact that $\gamma\in(0,1)$ and the assumptions on the continuity of $\dot{y}_{d}(t)$ and $f_{1}(\overline{x},t)+f_{0}(\underline{x},t)$ (see (F2) and (F3)). It is also worth noting that, as $f_{1}(\overline{x},t)$ and $f_{0}(\underline{x},t)$ describe the probability density  of TCLs in the ON and OFF states at the prescribed upper and lower temperature boundaries $\overline{x}$ and $\underline{x}$, respectively, it is impossible in practice that $f_{1}(\overline{x},t)+f_{0}(\underline{x},t)\rightarrow 0$ as $t\to +\infty$.
\end{remark}

\subsection{Finite-time input-to-state stability of the tracking error dynamics}\label{Sec: FTISS}
In this section, we assess the robust stability of the tracking error dynamics in the sense of FTISS, with $\Gamma$ as the input (disturbance). One of the main properties of the closed-loop system is stated below.
\begin{theorem}\label{Them: ftiss}
The power tracking error dynamics \eqref{error} under the control law given in \eqref{ideal-controller} are FTISS {w.r.t.} $\Gamma(t)$ for any $\gamma \in (0,1)$.
\end{theorem}
\begin{pf*}{Proof.} 
Consider a Lyapunov candidate of the form $V(e)=\frac{1}{2}e^{2}$.
The time derivative of $V$ along the trajectory of the tracking error dynamics~\eqref{error} is given by:
\begin{align*}
\dot{V} & =e\dot{e} 
  =e\left(-\frac{P}{\eta}k|e|^{\gamma}\mathrm{sgn}(e)+ \Gamma\right) 
   =-\frac{P}{\eta}k|e|^{1+\gamma}+e\Gamma 
   =-\frac{P}{\eta}k\left(\sqrt{2V}\right)^{1+\gamma}+e\Gamma 
  =-\frac{P}{\eta}k\left(2V\right)^{\frac{1+\gamma}{2}}+e\Gamma,
\end{align*}
%{\color{red} It should be that $\dot{e} = \left(-\frac{P}{\eta}k|e|^{\gamma}\mathrm{sgn}(e)+\Gamma\right)$. The following deductions are right, but the second line in the original file is wrong.}
which implies that
\begin{equation}
DV(e)\cdot f(e,\Gamma)\leq-\frac{P}{\eta}k(2V)^{\frac{1+\gamma}{2}}+ |e||\Gamma| \label{DV}
\end{equation}
with $f(e,\Gamma):=-\frac{P}{\eta}k|e(t)|^{\gamma}\mathrm{sgn}(e(t))  +\Gamma(t)$.

Let $C_{0}\in(0,k)$ be a constant. Then, for any $|e|\geq \left(\frac{\eta}{PC_{0}}|\Gamma|\right)^{\frac{1}{\gamma}}$, i.e., $|\Gamma |\leq\frac{P}{\eta}C_{0}|e|^{\gamma}$,  we deduce by \eqref{DV} that
\begin{align*}
DV(e)\cdot f(e,\Gamma)  \leq-\frac{P}{\eta}k(2V)^{\frac{1+\gamma}{2}}+\frac{P}{\eta}C_{0}|e|^{1+\gamma} 
   =-\frac{P}{\eta}k(2V)^{\frac{1+\gamma}{2}}+\frac{P}{\eta}C_{0}(2V)^{\frac{1+\gamma}{2}} 
  =-\frac{P}{\eta}(k-C_{0})2^{\frac{1+\gamma}{2}} V^{\frac{1+\gamma}{2}}.
\end{align*}

Note that {$\frac{P}{\eta}(k-C_{0})2^{\frac{1+\gamma}{2}}>0$, $\frac{1+\gamma}{2}\in(\frac{1}{2},1)$, and that
$\chi(s):=(\frac{\eta}{PC_{0}}s)^{\frac{1}{\gamma}}$} is a $\mathcal{K}$-function w.r.t. $s\in\mathbb{R}_{\geq 0}$. The FTISS of system~\eqref{error} is then guaranteed by Lemma~\ref{LemmaLyapunov}.
$\hfill\blacksquare$
\end{pf*}

\subsection{Properties of the governing PDEs}\label{Sec: Properties of the governing PDEs}
In practice, we can assume that the number of TCLs in a population remains unchanged within a specific DR control period. Therefore, the mass conservation property of the solutions to the system \eqref{eq: pdes}-\eqref{eq: IC} should be verified under the imposed boundary conditions, thereby conforming the compliance of the mathematical model with the imposed condition. Moreover, non-negativeness  of the solutions is also required.
\begin{theorem}[Mass conservation property]\label{Prop. conservation}
 The solution to the initial-boundary value problem (IBVP) \eqref{eq: pdes}-\eqref{eq: IC} is conservative in the sense that
\begin{align}\label{eq: conservativity}
  \int_{x_L}^{\overline{x}(t)}f_0(x,t)\diff{x} + \int_{\underline{x}(t)}^{x_H}f_1(x,t)\diff{x} = 1 \ \ \forall t \in \mathbb{R}_{\geq 0},
\end{align}
provided that
\begin{align}\label{eq: conservativity IC}
&\int_{x_{L}}^{\underline{x}(0)}f_{0}^{a0}(x)\diff{x}+\int_{\underline{x}(0)}^{\overline{x}(0)}f_{0}^{b0}(x)\diff{x} +\int_{\underline{x}(0)}^{\overline{x}(0)}f_{1}^{b0}(x)\diff{x}+\int_{\overline{x}(0)}^{x_{H}}f_{1}^{c0}(x)\diff{x} =1.
\end{align}
%\begin{align}
%& \int_{x_L}^{\underline{x}(t)}v(x,t)\diff{x}+\int_{\underline{x}(t)}^{\overline{x} (t)}v(x,t)\diff{x}
% + \int_{\underline{x}(t)}^{\overline{x} (t)} w(x,t)\diff{x}+\int_{\overline{x}(t)}^{x_H}w(x,t)\diff{x}\notag\\
%=&\int_{x_L}^{\underline{x}(0)}v_{a0}(x)\diff{x}+\int_{\underline{x}(0)}^{\overline{x} (0)}v_{b0}(x)\diff{x}
% + \int_{\underline{x}(0)}^{\overline{x} (0)} w_{b0}(x)\diff{x}+\int_{\overline{x}(0)}^{x_H}w_{c0}(x)\diff{x},\forall t\in \mathbb{R}_{\geq 0}.
%\end{align}
\end{theorem}

\begin{pf*}{Proof.} 
Using  \eqref{Pa}, \eqref{Pb1}, \eqref{BC1}, \eqref{BC2}, and \eqref{BC3}, and noting (U) and (F2),  we have
\begin{align}
  &\frac{\text{d}}{\diff{t}}\left(\int_{x_{L}}^{\overline{x}(t)}f_{0}(x,t)\diff{x}\right) \notag\\
= & \frac{\diff{}}{\diff{t}}\left(\int_{x_{L}}^{\underline{x}(t)}f_{0}(x,t)\diff{x}+\int_{\underline{x}(t)}^{\overline{x}(t)}f_{0}(x,t)\diff{x}\right)\notag\\
= & \int_{x_{L}}^{\underline{x}(t)}\partial_{t}f_{0}(x,t)\diff{x}+f_{0}(\underline{x}(t),t)\dot{\underline{x}}(t)+\int_{\underline{x}(t)}^{\overline{x}(t)}\partial_{t}f_{0}(x,t)\diff{x}  +f_{0}(\overline{x}(t),t)\dot{\overline{x}}(t)-f_{0}(\underline{x}(t),t)\dot{\underline{x}}(t)\notag\\
= & \int_{x_{L}}^{\underline{x}(t)}\partial_{x}\!\left(\frac{\sigma^{2}}{2}\partial_{x}f_{0}(x,t)-(\alpha_{0}(x,t)-u(t))f_{0}(x,t)\right)\diff{x}  +\int_{\underline{x}(t)}^{\overline{x}(t)} \partial_{x}\!\left(\frac{\sigma^{2}}{2}\partial_{x}f_{0}(x,t)-(\alpha_0(x,t)-u(t))f_{0}(x,t)\right) \diff{x}\notag\\
 &-\int_{\underline{x}(t)}^{\overline{x}(t)}  g(f_{0},f_{1}) \diff{x}\notag\\
= & \left(\frac{\sigma^{2}}{2}\partial_{x}f_{0}(x,t)-(\alpha_0(x,t)-u(t))f_{0}(x,t)\right) \bigg|_{x_{L}^{+}}^{\underline{x}^{-}(t)}\  +\left(\frac{\sigma^{2}}{2}\partial_{x}f_{0}(x,t)-(\alpha_0(x,t)-u(t))f_{0}(x,t)\right) \bigg|_{\underline{x}^{+}(t)}^{\overline{x}^{-}(t)}\notag\\
 & -\int_{\underline{x}(t)}^{\overline{x}(t)}g(f_{0},f_{1})\diff{x}\notag\\
= & \frac{\sigma^{2}}{2}\partial_{x}f_{0}(\underline{x}^{-}(t),t)-(\alpha_{0}(\underline{x}(t))-u(t))f_{0}(\underline{x}(t),t)-0  +\frac{\sigma^{2}}{2}\partial_{x}f_{0}(\overline{x}^{-}(t),t)-(\alpha_{0}(\overline{x}(t))-u(t))f_{0}(\overline{x}(t),t)\notag\\
 & -\left(\frac{\sigma^{2}}{2}\partial_{x}f_{0}(\underline{x}^{+}(t),t)-(\alpha_{0}(\underline{x}(t))-u(t))f_{0}(\underline{x}(t),t)\right)  -\int_{\underline{x}(t)}^{\overline{x}(t)}g(f_{0},f_{1})\diff{x}\notag\\
= &
\frac{\sigma^{2}}{2}(\partial_{x}f_{0}(\underline{x}^{-}(t),t)-\partial_{x}f_{0}(\underline{x}^{+}(t),t)) +\frac{\sigma^{2}}{2}\partial_{x}f_{0}(\overline{x}^{-}(t),t)-\int_{\underline{x}(t)}^{\overline{x}(t)}g(f_{0},f_{1})\diff{x}\notag\\
= & \frac{\sigma^{2}}{2}\partial_{x}f_{1}(\underline{x}^{+},t)+\frac{\sigma^{2}}{2}\partial_{x}f_{0}(\overline{x}^{-}(t),t)-\int_{\underline{x}(t)}^{\overline{x}(t)}g(f_{0},f_{1})\diff{x}.\label{2}
\end{align}
Similarly, we infer from \eqref{Pb2}, \eqref{Pc}, \eqref{BC4}, \eqref{BC5},  \eqref{BC6},  (U) and (F3) that
\begin{align}
  \frac{\diff{}}{\diff{t}}\left(\int_{\underline{x}(t)}^{x_H}f_1(x,t)\diff{x}\right) 
 =& \frac{\text{d}}{\text{d}t}\left( \int_{\underline{x}(t)}^{\overline{x} (t)} f_1(x,t)\diff{x}+\int_{\overline{x}(t)}^{x_H}f_1(x,t)\diff{x}  \right) \notag\\
 =& -\frac{\sigma^2}{2} \partial_{x}f_1(\underline{x}^+,t)
    -\frac{\sigma^2}{2} \partial_{x}f_0(\overline{x}^- (t),t)
    +\int_{\underline{x}(t)}^{\overline{x} (t)}g(f_0,f_1) \diff{x}.\label{3}
\end{align}
By \eqref{2} and \eqref{3}, we obtain
\begin{align*}
  \frac{\text{d}}{\text{d}t}\left( \int_{x_L}^{\overline{x}(t)}f_0(x,t)\diff{x} + \int_{\underline{x}(t)}^{x_H}f_1(x,t)\diff{x} \right)
 = 0 \ \ \forall t \in \mathbb{R}_{\geq 0},
\end{align*}
which along with \eqref{eq: conservativity IC} implies \eqref{eq: conservativity}.
$\hfill\blacksquare$
\end{pf*}
\begin{theorem} [Non-negativeness] \label{Prop-nonnegative}
The following statements hold true for the solution to IBVP~\eqref{eq: pdes}-\eqref{eq: IC}:
\begin{enumerate}
 \item[(i)] $f_0(x,t)\geq 0$  for all $x\in [x_L,\overline{x}(t)]$ and all $t\in \mathbb{R}_{\geq 0}$;
 \item[(ii)] $f_1(x,t)\geq 0$  for all $x\in [ \underline{x}(t),x_H]$ and all $t\in \mathbb{R}_{\geq 0}$;
  \item[(iii)] $f_0(\underline{x}(t),t)+f_1(\overline{x}(t),t)>0$ for all $t\in \mathbb{R}_{> 0}$.
\end{enumerate}
\end{theorem}

The proof of this theorem is provided in Appendix.

\section{Experimental Validation}\label{Sec: experiment}
In this section, we present simulation results to demonstrate the effectiveness of the proposed control scheme. Note that the control law given in \eqref{ideal-controller} is derived from the coupled Fokker-Planck equations, which assume a population of an infinite number of TCLs. As the number of TCLs in a real-world TCL population is always finite, and considering the fact that the larger the population size, the more accurate the PDE model, we present a comparative study of two heterogeneous populations with 1,000 and 100,000 TCLs.

\subsection{Simulation setup}
A numerical simulation is conducted to validate the proposed control scheme and evaluate its performance. Table~1 lists the physical parameters of the AC units utilized in the simulation, which are the same as those in \cite{Callaway:2009}. The thermal resistances and thermal capacitances are random variables following a log-normal distribution with average mean values of {$2$}~$^{\circ}$C/kW and {$10$}~kWh/$^{\circ}$C, respectively. The level of heterogeneity is parameterized by the standard deviation $\sigma_p$. In our experiment, the initial temperatures of the AC units are uniformly distributed around the initial set-point $x_{sp}^0 = 20$~$^{\circ}$C over the deadband, and initially $40\%$ of the AC units are set randomly in ``ON''-state. This setting causes the population to begin running from an almost steady state.

\begin{center}Table 1: Simulation parameter
\end{center} 
 
\begin{table}[ht] 
\centering
\begin{tabular}{|l|l|c|c|c|c|}
\hline
\textbf{Parameter} & \textbf{Description (Unit)} & \textbf{Value}\\
\hline
$R$        & average thermal resistance ($^\circ$C/kW) & 2\\
$C$        & average thermal capacitance (kWh/$^\circ$C) & 10\\
$P$        & electric power (kW) & 14\\
$\eta$     & load efficiency &  2.5 \\
$x_{sp}^0$ & initial temperature set-point ($^\circ$C) & 20\\
$\delta$   & temperature deadband width ($^\circ$C) & 0.5\\
$\sigma_p$ & standard deviation of lognormal distributions & 0.2\\
\hline
$p_f$      & forced switch probability per hour (\%) & 3\\
$t_{ci}$   & control interval (second) & 30 \\
$t_{\mathrm{lock}}$ & locked time of each TCL (minute) & 6\\
\hline
\end{tabular}
\end{table}

\begin{figure}[!htbp]
    \centering
    \captionsetup{justification=centering}
    \includegraphics[width=0.52\textwidth]{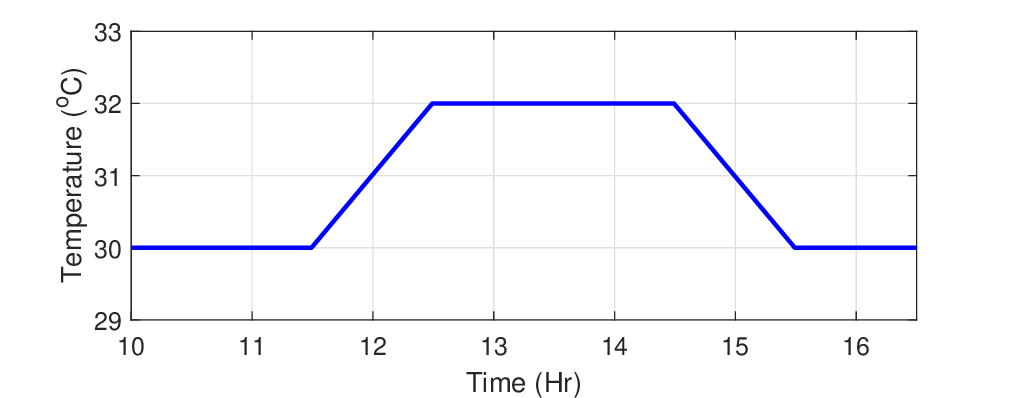}
    \caption{Ambient temperature.}
  \label{fig:outside}
\end{figure}

The disturbances brought into the system come mainly from the following three sources. First, all AC units operate under the same varying outside temperature, as depicted in Fig.~\ref{fig:outside}, which rises from $30$$^{\circ}$C at 11:30 to $23$$^{\circ}$C at 12:30 and then drops back from 14:30 to 15:30. Second, a forced random switch mechanism is added to desynchronize AC operations. The number of forced interrupts per hour can be adjusted through the hyper-parameter $p_f$. Moreover, a safe border distance of {$5\%$} of the deadband width is incorporated to prevent forced switches from happening when an AC is around $\overline{x}(t)$ and in ``ON'' state or around $\underline{x}(t)$ and in ``OFF'' state. Finally, because frequent switching leads to reduced energy efficiency and more rapid compressor wear out, a lockout time, $t_{\mathrm{lock}}$, is included for each AC. Thus, an AC unit remains inactive to the control signals when it is locked.

\begin{figure}[!htbp]
    \centering
    \captionsetup{justification=centering}
    \includegraphics[width=0.52\textwidth]{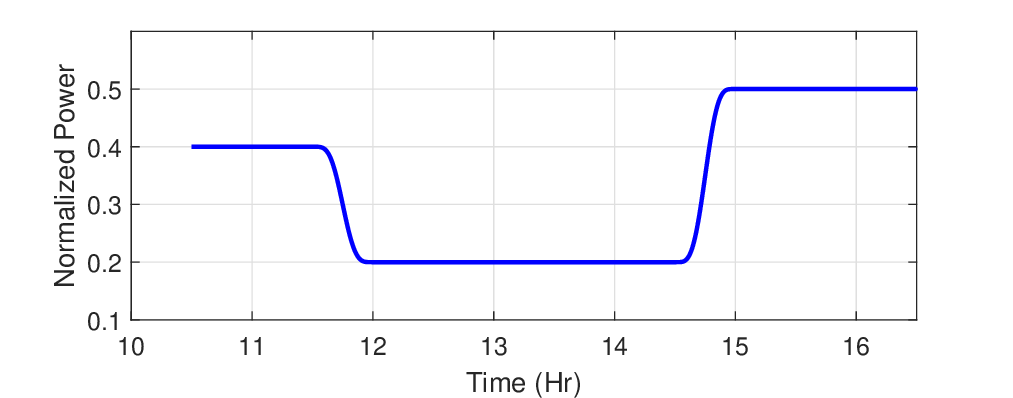}
    \caption{Desired power profile.}
  \label{fig:ReferencePower}
\end{figure}

The reference power is a predefined curve, as shown in Fig.~\ref{fig:ReferencePower}. From 10:30 to 11:30, the normalized desired power is maintained constant at {$0.4$}. From 11:30 to 12:00, the reference power drops to $0.2$ and keeps constant for the following two and a half hours. From 14:30, the desired power rises to $0.5$ in 30 minutes and remains constant until 16:30. During the rising and dropping phases, the desired power is specified by a smooth polynomial with the endpoint constraints given below:
\begin{align}\label{eqn:curve}
  y_{d}(t)= \left(y_{d}(t_{f}) -y_{d}(t_{i})\right)\tau^{5}(t)\sum_{l=0}^{4}a_{l}\tau^{l}(t), t\in[t_{i},t_{f}],
\end{align}
\begin{align}\label{eqn:constraint}
  {\dot{y}}_{d}(t_{i})={\dot{y}}_{d}(t_{f})={\ddot{y}}_{d}(t_{i}) ={\ddot{y}}_{d}(t_{i}) ={\overset{\ldots}{y}}_{d}(t_{i})={\overset{\ldots}{y}}_{d}(t_{f})=0,
\end{align}
where $\tau(t)=(t-t_{i})/(t_{f}-t_{i})$. By a direct computation, the coefficients can be determined as follows:
\begin{align*}
  a_{0}=126,\,a_{1}=420,\,a_{2}=540,\,a_{3}=315,\,\mathrm{and}\ a_{4}=70.
\end{align*}

In the simulation, the control signal is updated every 30 seconds ($t_{ci}$ in Table~\ref{tab:table1}). The control signal that every AC receives is the set-point variation rate. Each AC computes then its set-point temperature offset for the next control interval starting from $t_{k}$. To compute the denominator of the controller given in \eqref{ideal-controller}, a mid-point rectangular method with a temperature bin width $\delta_x$ is used to estimate $f_{1}(\overline{x}(t_{k}),t_{k})$ and $f_{0}(\underline{x}(t_{k}),t_{k})$. The percentage of ACs falling in the rectangular region is used as $f_{1}(\overline{x}(t_{k}),t_{k}) \times \delta_x$ or $f_{0}(\underline{x}(t_{k}),t_{k}) \times \delta_x$. In general, $\delta_x$ should not be too large because the underlying system has complex nonlinear dynamics. On the other hand, considering the limited number of ACs involved in the simulation, the bin width $\delta_x$ should not be too small, which may introduce larger biases. In our implementation, histogram bin widths of $0.008~^\circ$C, $0.004~^\circ$C, and $0.002~^\circ$C are used, which are reasonable and provide reliable estimations of $f_{1}(\overline{x}(t_{k}),t_{k})$ and $f_{0}(\underline{x}(t_{k}),t_{k})$.

\begin{table}[!htbp]
  \begin{center}
Table 2: Tracking performance of 10 episodes for the population with 1,000 TCLs\\
~\\
  \begin{tabular}{|c|c|c|c|c|c|c|c|}
    \hline
    \textbf{Episode}      & 1  & 2 & 3 & 4 & 5 \\ \hline
    \textbf{RMSE~($\%$)} & 0.948 & 0.923 & 0.844 & 0.834 & 0.935 \\ \hline
    \textbf{Episode}    & 6  & 7 & 8 &  9 & 10\\ \hline
    \textbf{RMSE~($\%$)} & 0.880 & 0.923 & 0.890 & 0.925 & 0.861\\ \hline
  \end{tabular}
  \label{tab:rmse}
  \end{center}
\end{table}

\subsection{Simulation results}

First, we present the test results for the population with 1,000 TCLs. The control cycle lasts for 6~hours, from 10:30 to 16:30. The test is performed continuously for $10$ episodes, and the tracking performance is measured by the root mean square error (RMSE), as reported in Table~2. In the test, the controller parameters in \eqref{ideal-controller} are set to be $k = 8$ and $\gamma = 0.5$, respectively. The final result shows that the mean RMSE for this setting is $0.896\%$, and the standard deviation (STD) of the dRMSEs is $0.040\%$.

\begin{figure}[!htbp]
     \centering
     \begin{subfigure} 
          
         \includegraphics[width=0.52\textwidth]{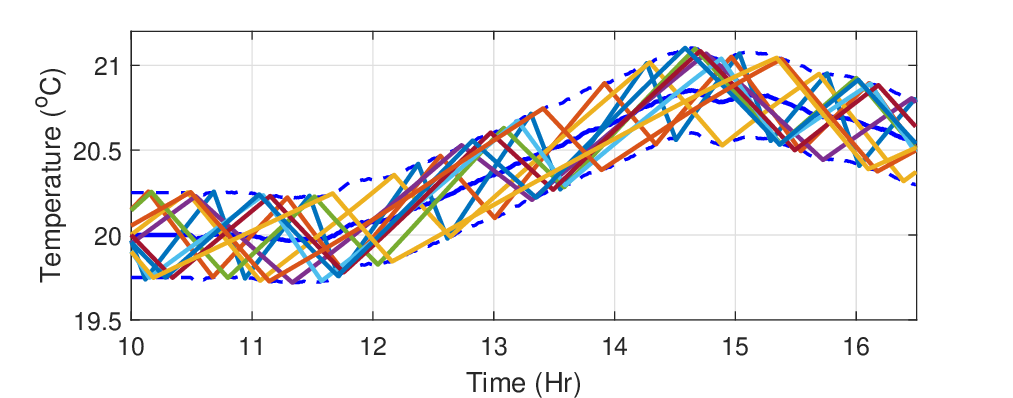}\\
       (a) tracking performance
                  \label{fig:TrackingCaseI}
      
         \includegraphics[width=0.52\textwidth]{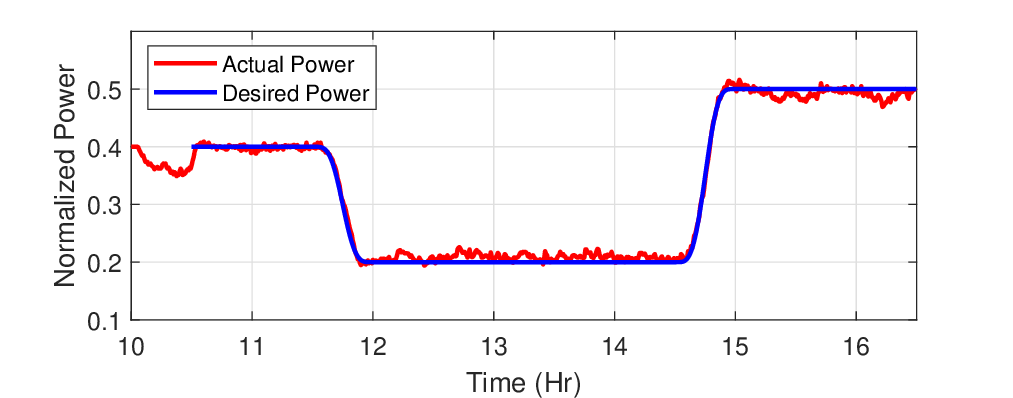}\\
         (b) temperature trajectories of $10$ ACs
                 \label{fig:TrajectoryCaseI}
     
         \includegraphics[width=0.52\textwidth]{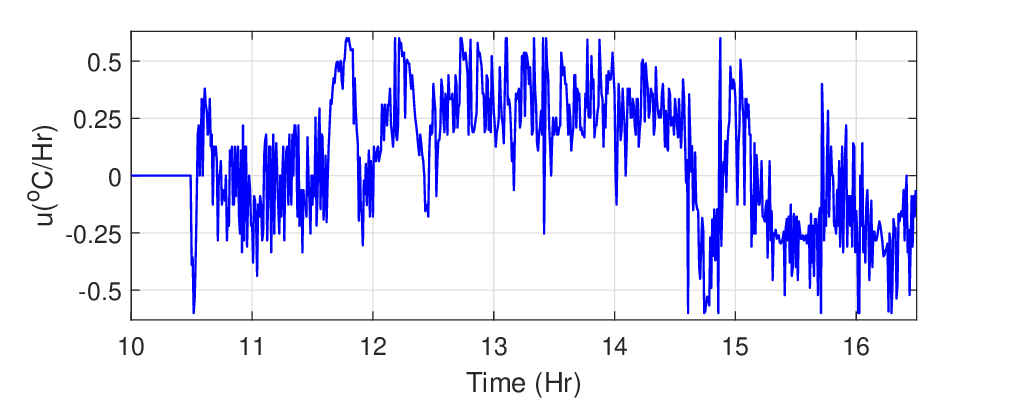}\\
      (c) set-point variation rate
         \label{fig:OffsetCaseI}
     \end{subfigure}
     \caption{Control performance for a population of 1,000 TCLs}
     \label{fig:CaseI}
\end{figure}

Fig.~\ref{fig:CaseI} shows a sample of the control results corresponding to the episode with an RMSE of $0.948\%$. It can be seen from Fig.~5a that the proposed control strategy is effective. The temperature evolution of $10$ randomly selected ACs in the population is presented in Fig.~5b. It can be observed that all of them, unless forced switches occur, operate smoothly inside the deadband between the turning on and turning off points. Fig.~5c shows the control signal generated during this episode. During the first 30~minutes (from 10:00 to 10:30), the controller is inactive, and the system operates in an open-loop mode.

\begin{figure}[htbp]
     \centering
     
     \begin{subfigure}
     
         \includegraphics[width=0.52\textwidth]{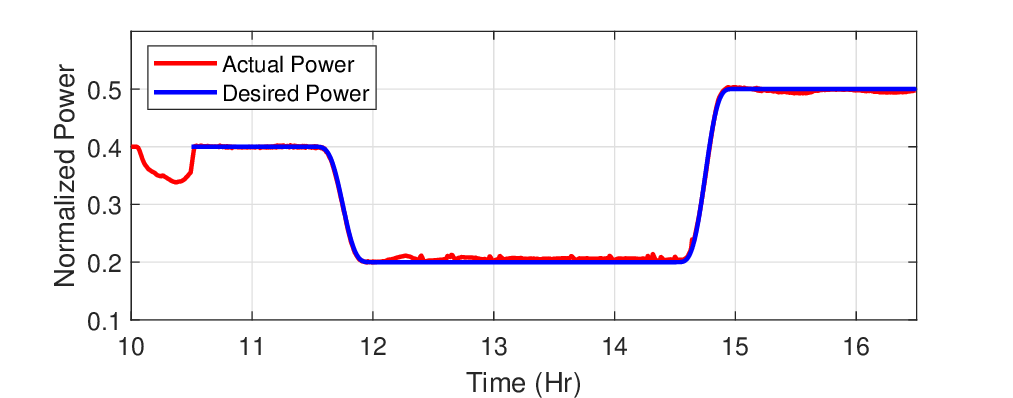}\\
         (a) tracking performance
         
      \includegraphics[width=0.52\textwidth]{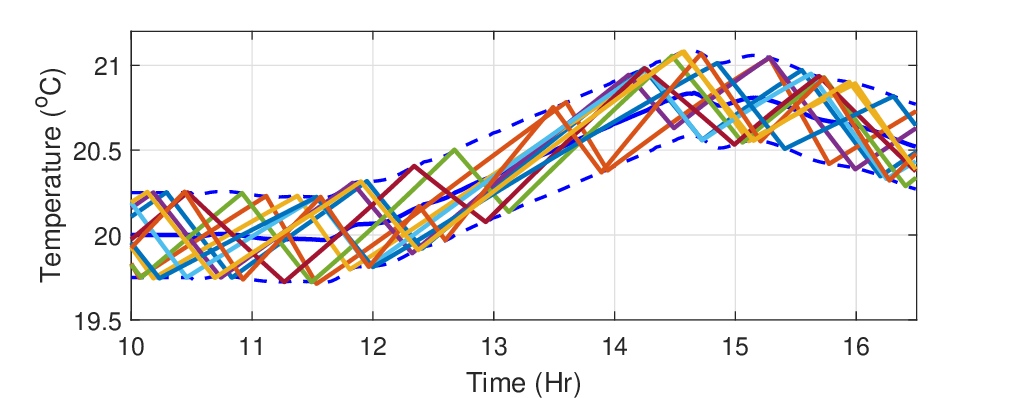}\\
      (b) temperature trajectories of $10$ ACs
      
         \includegraphics[width=0.52\textwidth]{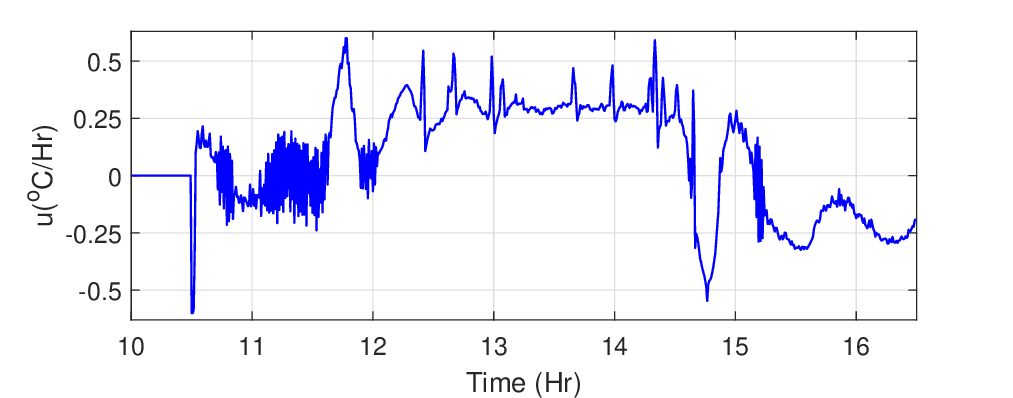}\\
         (c) set-point variation rate
\end{subfigure}
     \caption{Control performance for a population of 100,000 TCLs}
     \label{fig:CaseII}
\end{figure}

When the number of ACs increases, the model of the coupled Fokker-Planck equations becomes more accurate. To evaluate the effectiveness of the proposed control strategy, tracking control performance is examined for a population of 100,000 ACs. The RMSE values for 10 continuous tests are shown in Table~\ref{tab:rmse2}, which gives a mean RMSE of $0.497\%$ and an STD of $0.004\%$. In this test, $k=15$ and $\gamma=0.5$ are used. Fig.~\ref{fig:CaseII} illustrates one of the control samples corresponding to the episode with an RMSE of $0.505\%$. The normalized power consumption is shown in Fig.~6a, and the temperature evolutions of $10$ ACs are shown in Fig.~6b. The control signal is shown in Fig.~6c.
\begin{table}[!htbp]
  \begin{center}
  \caption{Tracking performance of 10 episodes for the population with 100,000 TCLs}
  \begin{tabular}{|c|c|c|c|c|c|c|c|}
    \hline
    \textbf{Episode}      & 1  & 2 & 3 & 4 & 5 \\ \hline
    \textbf{RMSE~($\%$)} & 0.505 & 0.500 & 0.496 & 0.491 & 0.490 \\ \hline
    \textbf{Episode}    & 6  & 7 & 8 &  9 & 10\\ \hline
    \textbf{RMSE~($\%$)} & 0.497 & 0.499 & 0.495 & 0.498 & 0.500\\ \hline
  \end{tabular}
  \label{tab:rmse2}
  \end{center}
\end{table}

The results of the comparative study show clearly that the tracking control system performs better for the population of larger size with smaller RMSE, smoother power trajectory, and less ``noisy'' control signals. This is consistent with the nature of the PDE model on which the proposed control scheme is based. Nevertheless, the performance is not significantly degraded for a population with a significantly smaller size. This demonstrates the robustness and potential applicability of the developed control strategy to practical systems.

\section{Conclusion}\label{Sec: Conclusion}
In this work, we have developed a strategy for the power tracking control of heterogeneous TCL populations. The control scheme can ensure a robust performance in the presence of modeling uncertainties in the sense of FTISS and requires measuring the states of the system only on the end-points of the deadband. The simulation results provided encouraging evidence that the proposed control approach is highly effective. From a practical application viewpoint, we can consider in our future work other types of devices, such as battery charging systems, and other demand-response tasks, such as frequency regulation or transaction control \cite{transaction1,transaction2, transaction3}. Control of populations of TCLs described by the second-order equivalent thermal parameters model \cite{liu2019interval,cheng2020day} may also be a research direction worthy of exploration.

\section*{Appendix:  Proof of Theorem~5}

We first prove statement~(i). Given any $T>0$, it suffices to show that  $f_0\geq 0$ over $[x_L,\overline{x}(t )]\times [0,T]$ for all $t\in [0,T]$.

Indeed, the transformations of variable  $y:=\frac{x-x_L}{\overline{x}-x_L}:=\frac{x-x_L}{h}$ and $f_0(x,t)= f_0(yh+x_L,t):=\tilde{f}_0(y,t)$   yield
\begin{align*}
\partial_{x}f_{0} & =\frac{1}{h}\partial_{y}\tilde{f}_{0},\notag\\
\partial_{xx}f_{0}& =\frac{1}{h^{2}}\partial_{yy}\tilde{f}_{0},\partial_{t}f_{0} =\partial_{t}\tilde{f}_{0}+\partial_{y}\tilde{f}_{0}\frac{\partial y}{\partial t}  =\partial_{t}\tilde{f}_{0}-(x-x_{L})\frac{\dot{\overline{x}}}{h^{2}}\partial_{y}\tilde{f}_{0} =\partial_{t}\tilde{f}_{0}-\frac{1}{h}yu\partial_{y}\tilde{f}_{0}.
\end{align*}
Note that
\begin{align*}
  & x\in [x_L,\underline{x} ] \Leftrightarrow y\in [0,1] , \notag\\
 &0<\delta_0\leq h(t)\leq x_H-x_L,\forall t\in [0,T].
\end{align*}
% Note that
% \begin{align*} x\in [x_L,\underline{x} ] \Leftrightarrow y\in [0,1] , \notag\\
% 0<\delta_0\leq h(t)\leq x_H-x_L,\forall t\in [0,T].
% \end{align*}

The PDEs \eqref{Pa} and  \eqref{Pb1} are equivalent to
\begin{subequations}\label{tilde-v}
\begin{align}
 \partial_{t}\tilde{f}_{0}-\frac{1}{h}
\left(\frac{\sigma^{2}}{2h}\partial_{yy}\tilde{f}_{0}+\left((1+y)u-\tilde{\alpha}_{0}\right)\partial_{y}\tilde{f}_{0}
   -\tilde{\alpha}_{0y}\tilde{f}_{0}\right)  
=& 0, \ \ \forall y\in\left(0,z(t)\right), \forall t\in(0,T],\\
 \partial_{t}\tilde{f}_{0}-g(\tilde{f}_{0},\tilde{f}_{1})-\frac{1}{h}
 \left(\frac{\sigma^{2}}{2h}\partial_{yy}\tilde{f}_{0}+\left((1+y)u-\tilde{\alpha}_{0}\right)\partial_{y}\tilde{f}_{0} -\tilde{\alpha}_{0y}\tilde{f}_{0}\right)
  =&0, \ \ \forall y\in\left(z(t),1\right), \forall t\in(0,T],
\end{align}
\end{subequations}
respectively, where $\tilde{\alpha}_0(y,t):=\alpha_0(yh(t) +x_L,t)$, $f_1(x,t)=  f_1(yh+x_L,t):=\tilde{f}_1(y,t)$, and $z(t):=1-\frac{\delta_0}{h(t)}$.

Note that \eqref{eq: BC}  is  equivalent to \eqref{eq: BC-pf}, and \eqref{BC1}, \eqref{BC2},  and \eqref{BC3} become
\begin{subequations}\label{tilde-v-BCs}
\begin{align}
 \frac{\sigma^{2}}{2}\partial_{y}\tilde{f}_{0}(0^{+},t)-
(\tilde{\alpha}_{0}(0^{+},t)-u(t))h(t)\tilde{f}_{0}(0^{+},t) =&0,  \    \forall t\in(0,T], \\
 \partial_{y}\tilde{f}_{0}\left(z^{-}(t),t\right)-\partial_{y}\tilde{f}_{0}\left(z^{+}(t),t\right) =&\sigma_{0}(t),\ \forall t\in(0,T], \\
 \tilde{f}_{0}(1^{-},t) =&0 ,\  \forall t\in(0,T],
\end{align}
\end{subequations}
where, for the given solution $f_1$, $\sigma_0(t):=\frac{\sigma^2}{2}\partial_{x}f_1 (\underline{x}^+(t),t)$ is a well-defined function {w.r.t.} $t$, and $\sigma_0(t)>0$ for all $t\in [0,T]$ owing to (F3) and \eqref{BC8}.

The initial data of $\tilde{f}_0$  over the domain $\left [0, z(t) \right]$ and $ \left [ z(t),1\right]$ are given by
\begin{align*}
  \tilde{f}_0^{a0}(y) :=f_0^{a0}(yh(0) +x_L)\geq 0,
\end{align*}
and
\begin{align*}
\tilde{f}_0^{b0}(y) :=f_0^{b0}(yh(0) +x_L)\geq 0,
\end{align*}
respectively.

% Due to the continuity of $v$, for any constant $T>t_1$, there must be a point $(x_0,t_0)\in [ x_L,\overline{x}(t_0)]$ such that $v$ attains its negative minimum over the domain $[ x_L,\overline{x}(t)]\times [0,T]$

Let $\phi(y):=\e^{m(y-\frac{1}{2})^2}$ and $\tilde{f}_0:=\phi \e^{\gamma t}\hat{f}_0$ with $m>0$ and $\gamma>0$ being constants that will be chosen later. Then  \eqref{tilde-v} and \eqref{tilde-v-BCs} lead to
\begin{subequations}
\begin{align}
 \partial_{t}\hat{f}_{0}-\frac{\sigma^{2}}{2h^{2}}\partial_{yy}\hat{f}_{0}+ \mathcal{B}(y,t)\partial_{y}\hat{f}_{0}+\mathcal{C}(y,t)\hat{f}_{0}
  =&0, \forall y\in\left(0,z(t)\right), \forall t\in(0,T],\label{hat-v-a} \\
 \partial_{t}\hat{f}_{0}-\frac{\sigma^{2}}{2h^{2}}\partial_{yy}\hat{f}_{0}+\mathcal{B}(y,t)\partial_{y}\hat{f}_{0} +\mathcal{C}(y,t)\hat{f}_{0} + \frac{e^{-\gamma t}}{\phi(y)}g(\tilde{f}_{0},\tilde{f}_{1})  
 =&0, \forall y\in\left(z(t),1\right), \forall t\in(0,T],\label{hat-v-b}\\
 \frac{\sigma^{2}}{2}\partial_{y}\hat{f}_{0}(0^{+},t)-k(t)\hat{f}_{0}(0^{+},t) =&0,   \forall t\in(0,T],\label{hat-v-BC1}\\
 \partial_{y}\hat{f}_{0}\left(z^{-}(t),t\right)-\partial_{y}\hat{f}_{0}\left(z^{+}(t),t\right) =&\hat{\sigma}_{0}(t),   \forall t\in(0,T],\label{hat-v-BC2}\\
 \hat{f}_{0}(1^{-},t) =&0,   \forall t\in(0,T], \label{hat-v-BC3}
\end{align}
%\begin{align}
% \partial_{t}\hat{f}_0- \frac{\sigma^2}{2h^2}\partial_{yy}\hat{f}_{0}+\mathcal{B}(y,t)\partial_{y}\hat{f}_0
%+\mathcal{C}(y,t) \hat{f}_0 =&0,\forall y\in \left (0,z(t)\right) ,\forall t\in (0,T], \label{hat-v-a} \\
% \partial_{t}\hat{f}_0- \frac{\sigma^2}{2h^2}\partial_{yy}\hat{f}_{0}+\mathcal{B}(y,t)\partial_{y}\hat{f}_0
%+\mathcal{C}(y,t) \hat{f}_0 +\frac{\e^{-\gamma t}}{\phi(y)}g(\tilde{f}_0,\tilde{f}_1)=&0,\forall y\in \left ( z(t),1\right) ,\forall t\in (0,T], \label{hat-v-b}\\
%  \frac{\sigma^2}{2}\partial_{y}\hat{f}_0(0^+,t)-k(t)\hat{f}_0(0^+,t) =&0,  \forall t\in (0,T],\label{hat-v-BC1}\\
% \partial_{y}\hat{f}_0\left( z  ^-(t),t\right)-\partial_{y}\hat{f}_0\left(  z^+(t),t\right)=&\hat{\sigma}_0(t),
%    \forall t\in (0,T],\label{hat-v-BC2} \\
%\hat{f}_0(1^-,t) =&0,  \forall t\in (0,T], \label{hat-v-BC3}
%%\hat{v}(y,0)=&\hat{v}_0(y),\forall y\in (0,1),\label{hat-v-d}
%\end{align}
\end{subequations}
where
\begin{align*}
 \mathcal{B}(y,t):=& -\frac{1}{h}\left(\frac{\sigma^2}{2h}\frac{2\partial_{y}\phi}{\phi}+(1+y)u-\tilde{\alpha}_0 \right),\\
\mathcal{C}(y,t):=&\frac{1}{h}\left(
\gamma- \frac{\sigma^2}{2h}\frac{\partial_{yy}\phi}{\phi}-\frac{\partial_{y}\phi}{\phi}\left((1+y)u
      -\tilde{\alpha}_0 \right)+\tilde{\alpha}_{0y}\right),  \\
k(t):=&  \frac{m\sigma^2}{2} +(\tilde{\alpha}_0(0^+,t)-u(t))h(t),\\
\hat{\sigma}_0(t):=&\frac{\e^{-\gamma t}}{\phi  (1)}\sigma_0(t).
%\hat{v}_0(y):=&\frac{\tilde{v}_0(y)}{\phi  (y)}.
\end{align*}

The initial data for the $\hat{f}_0$-system over the domain $ \left [0, z(t) \right]$ and $ \left [ z(t),1\right]$ are given by
\begin{align} \label{hat-v-IC}
  \hat{f}_0^{a0}(y):=\frac{\tilde{f}_0^{a0}(y)}{\phi  (y)}\geq 0\ \ \text{and}\ \
  \hat{f}_0^{b0}(y):=\frac{\tilde{f}_0^{b0}(y)}{\phi  (y)}\geq 0,
\end{align}
respectively.

Note that $u,\tilde{\alpha}_0$, and $\tilde{\alpha}_{0y}$ are continuous in $[0,1]\times [0,T]$. Letting first $m$ and then $\gamma$ be sufficiently large, there must be positive constants $k_0$ and $c_0$ such that
\begin{align}
  k(t)\geq& k_0,  \forall t\in (0,T],\label{k0}\\
 \mathcal{C}(y,t) -1
\geq&  c_0 ,\forall (y,t)\in (0,1)\times(0,T].\label{c0}
\end{align}

To prove the non-negativeness property of $f_0$, it suffices to show that $\hat{f}_0\geq 0$ in $[0,1]\times [0,T]$. We now proceed with the proof by contradiction. Assume that there exists a point $(y_0,t_0)\in [0,1]\times [0,T]$ such that
\begin{align*}
  \hat{f}_0(y_0,t_0)=\min_{(y,t)\in [0,1]\times [0,T]}\hat{f}_0(y,t)<0.
\end{align*}
Considering \eqref{hat-v-BC3} and \eqref{hat-v-IC}, we have $y_0\neq 1$ and $t_0\in (0,T]$.

\emph{Case 1}: $y_0 \in \left (0,z(t_0)\right)$. At point $(y_0,t_0)$, it holds that
\begin{align*}
  \partial_{t}\hat{f}_0(y_0,t_0)\leq 0, \partial_{y}\hat{f}_{0}(y_0,t_0)= 0, \partial_{yy}\hat{f}_{0}(y_0,t_0)\geq 0.
\end{align*}
Then  \eqref{hat-v-a} and \eqref{c0} imply that
\begin{align*}
\begin{split}
 0>&\left(c_0+1\right)\hat{f}_0(y_0,t_0)\geq \partial_{t}\hat{f}_0(y_0,t_0)
   -\frac{\sigma^2}{2h^2(t_0)}\partial_{yy}\hat{f}_{0}(y_0,t_0) +\mathcal{B}(y_0,t_0)\partial_{y}\hat{f}_0(y_0,t_0)+\mathcal{C}(y_0,t_0 )\hat{f}_0(y_0,t_0)  
  = 0,
\end{split}
\end{align*}
which leads to a contradiction.

\emph{Case 2}: $y_0 \in \left (z(t_0),1\right)$. At the point $(y_0,t_0)$, it  also holds that
\begin{align*}
  \partial_{t}\hat{f}_0(y_0,t_0)\leq 0,   \partial_{y}\hat{f}_{0}(y_0,t_0)= 0, \partial_{yy}\hat{f}_{0}(y_0,t_0)\geq 0 .
\end{align*}

In addition, using the Mean Value Theorem, (G1), and (G2), we obtain:
\begin{align*}
\begin{split}
  g(\tilde{f}_0(y_0,t_0),\tilde{f}_1(y_0,t_0))  
 = g(0,\tilde{f}_1(y_0,t_0))+\tilde{f}_0(y_0,t_0)g_{s}(s,\tilde{f}_1(y_0,t_0))|_{s=\xi} 
\leq   |\tilde{f}_0(y_0,t_0)| ,
\end{split}
\end{align*}
where $\xi$ is between $0$ and $\tilde{f}_0(y_0,t_0)$.

It follows that
\begin{align}\label{equ.10}
 \frac{\e^{-\gamma t_0}}{\phi(y_0)}g(\tilde{f}_0(y_0,t_0),\tilde{f}_1(y_0,t_0))
 \leq & |\tilde{f}_0(y_0,t_0)|\frac{\e^{-\gamma t_0}}{\phi(y_0)}  
    =  -\hat{f}_0(y_0,t_0).
\end{align}

From \eqref{hat-v-b}, \eqref{c0}, and \eqref{equ.10}, we obtain:
\begin{align*}
\begin{split}
 0>&c_0\hat{f}_0(y_0,t_0)\notag\\
\geq & \left(\mathcal{C}(y_0,t_0 )-1\right)\hat{f}_0(y_0,t_0)\notag\\
\geq &
 \mathcal{C}(y_0,t_0 )\hat{f}_0(y_0,t_0) +\frac{\e^{-\gamma t_0}}{\phi(y_0)}g(\tilde{f}_0(y_0,t_0),\tilde{f}_1(y_0,t_0))\notag\\
\geq & \partial_{t}\hat{f}_0(y_0,t_0)- \frac{\sigma^2}{2h^2(t_0)}\partial_{yy}\hat{f}_{0}(y_0,t_0) +\mathcal{B}(y_0,t_0)\partial_{y}\hat{f}_0(y_0,t_0)+\mathcal{C}(y_0,t_0 )\hat{f}_0(y_0,t_0) +\frac{\e^{-\gamma t_0}}{\phi(y_0)}g(\tilde{f}_0(y_0,t_0),\tilde{f}_1(y_0,t_0))\notag\\
=& 0 ,
\end{split}
\end{align*}
which leads to a contradiction.

\emph{Case 3}: $y_0 =0$. It follows that $\partial_{y}\hat{f}_{0}(0^+,t_0)\geq 0$,
which, along with \eqref{hat-v-BC1} and \eqref{k0}, yields
\begin{align*}
\begin{split}
    0&<-k_0\hat{f}_0(0^+,t_0)\leq -k(t_0)\hat{f}_0(0^+,t_0)   \leq  \frac{\sigma^2}{2}\partial_{t}\hat{f}_0(0^+,t)-k(t_0)\hat{f}_0(0^+,t)= 0.
\end{split}
\end{align*}
We get a contradiction.

\emph{Case 4}: $ y_0 =1 $. It follows that $\partial_{y}\hat{f}_{0}(1^+,t_0)\leq 0$,
which along with \eqref{hat-v-BC1} and \eqref{k0} yields
\begin{align*}
\begin{split}
    0& <-k_0\hat{f}_0(0^+,t_0)\leq -k(t_0)\hat{f}_0 (0^+,t_0) \leq  \frac{\sigma^2}{2}\partial_{y}\hat{f}_0(0^+,t)-k(t_0)\hat{f}_0(0^+,t)= 0.
\end{split}
\end{align*}
We get a contradiction.

\emph{Case 5}:  $y_0=z(t_0) $. It follows that $\partial_{y}\hat{f}_{0}(z^-(t_0),t_0)\leq 0$ and $\partial_{y}\hat{f}_{0}(z^+(t_0),t_0)\geq 0$,
which along with \eqref{hat-v-BC2} and $\hat{\sigma}_0(t)>0$ yields
\begin{align*}
  0\geq \partial_{y}\hat{f}_{0}(z^-(t_0),t_0)-\partial_{y}\hat{f}_{0}(z^+(t_0),t_0)=\hat{\sigma}_0(t_0)>0,
\end{align*}
leading to a contradiction.

Because we always obtain a contradiction in each case, we have shown that $\hat{f}_0\geq 0$ over the domain $[0,1]\times [0,T]$, which implies the non-negativeness property of $f_0$ over the domain $[x_L,\overline{x}(t)]\times [0,T]$ for all $t\in [0,T]$ and all $T\in\mathbb{R}_{>0}$.

Because the proof of statement~(ii) can proceed in the same way as above, we omit the details of the proof.

Finally, suppose that statement~(iii) fails to be true; then, for any given $T\in \mathbb{R}_{>0}$ there must be a $t_0\in (0,T]$ such that
\begin{align*}
  f_0(\underline{x}(t_0) ,t_0)+f_1(\overline{x}(t_0),t_0)=0,
\end{align*}
which, along with the non-negativeness property of $f_0$ and $f_1$, implies that $f_0$ and $f_1$ attain their minima at $(\underline{x}(t_0) ,t_0)$ and $(\overline{x}(t_0),t_0)$, respectively. Then, using the same argument as that in \emph{Case 5}, we obtain a contradiction. Therefore, statement~(iii) holds true.
$\hfill\blacksquare$

%\begin{pf*}{Proof.}
%
%$\hfill\blacksquare$
%\end{pf*}

%
%\bibliographystyle{plain}
%\bibliography{References}

%%%%%%%%%%%%%%%%%%%%%%%%%%%%%%%%%%%%%%%%%%%%%%
\end{document}